\documentclass[10pt]{iopart}

\usepackage{iopams}  
\usepackage{graphicx}
\usepackage{hyperref}

\hypersetup{
    colorlinks=true,
    linkcolor=green,    % Color for internal links (e.g., table of contents)
    citecolor=green,    % Color for citation links
    urlcolor=green      % Color for URLs
}

\newcommand{\bm}[1]{\mathbf{#1}}
\newcommand{\avgt}[1]{\left<#1 \right>_t}
\newcommand{\avgy}[1]{\left<#1 \right>_y}
\newcommand{\genex}{\texttt{GENE-X} }

\begin{document}

\title[Turbulence and Transport in Spectral Gyrokinetic Simulations]{Turbulence and Transport in Spectrally Accelerated full-$f$ Gyrokinetic Simulations}

\author{B. J. Frei$^1$, P. Ulbl$^1$, C. Pitzal$^1$, W. Zholobenko$^1$, F. Jenko$^{1,2}$}

\address{$^1$ Max-Planck Institute for Plasma Physics, Boltzmannstr. 2, Garching, D-85748, Germany}
\address{$^2$ Institute for Fusion Studies, The University of Texas at Austin, Austin, TX 78712, USA}
\ead{baptiste.frei@ipp.mpg.de}
\vspace{10pt}
\begin{indented}
\item[]March 2025
\end{indented}

\begin{abstract}
We investigate edge and scrape-off layer (SOL) turbulence and transport using the spectrally accelerated full-$f$ gyrokinetic (GK) code \texttt{GENE-X}, recently introduced in [B. J. Frei et al., arXiv:2411.09232 (2024)]. Extending previous work on the TCV-X21 scenario, we show that the velocity-space spectral approach not only reproduces outboard midplane profiles but also captures key features of trapped electron mode (TEM)-driven turbulence and transport, including fluctuation spectra, turbulent fluxes, phase shifts, and power crossing the separatrix, in close agreement with grid-based results. This agreement remains robust when increasing spectral resolutions. We further analyze the radial force balance (accurately satisfied) and the structure of the radial electric fields and poloidal flows in the edge and SOL. Finally, we contrast our results with Braginskii-like fluid models, which inherently neglect TEMs. These results confirm the spectral full-$f$ \texttt{GENE-X} approach as an efficient and first-principles tool for predicting edge and SOL turbulence.
\end{abstract}

%
% Uncomment for keywords
%\vspace{2pc}
%\noindent{\it Keywords}: XXXXXX, YYYYYYYY, ZZZZZZZZZ
%
% Uncomment for Submitted to journal title message
%\submitto{\JPA}
%
% Uncomment if a separate title page is required
%\maketitle
% 
% For two-column output uncomment the next line and choose [10pt] rather than [12pt] in the \documentclass declaration
\ioptwocol

\section{Introduction}
A key goal of fusion research in magnetically confined plasmas is to predict turbulence-driven particle and heat transport across the edge and into the scrape-off layer (SOL) via first-principles simulations \cite{ikeda2007,zohm2013,litaudon2022}. The gyrokinetic (GK) \cite{brizard2007} and drift-reduced Braginskii \cite{zeiler1997} models are leading global approaches to simulate the nonlinear turbulent dynamics in these regions. These two approaches are full-$f$, meaning that no separation between equilibrium and fluctuations is assumed. Within the full-$f$ GK model \cite{brizard2007,frei2020}, the full gyrocenter distribution function is self-consistently evolved, including both collisionless and collisional (with an appropriate collision operator \cite{ulbl2023,frei2022}) physics. As a consequence, it can be considered as a high-fidelity approach where all the details of the gyrocenter distribution functions are taken into account. Although the full-$f$ GK model provides an accurate description of turbulence and transport, full-$f$ GK simulations, based on either Eulerian \cite{mandell2020,michels2021}, Particle-in-Cell \cite{hager2022,bottino2025}, and semi-Lagrangian \cite{grandgirard2007} schemes to discretize the velocity-space, remain extremely computationally demanding in particular in the edge and SOL regions. On the other hand, drift-reduced Braginskii (or Braginskii-like fluid) models \cite{zeiler1997,tamain2016,stegmeir2019,giacomin2021} are inherently less expensive since only the three lowest-order fluid moments (density, velocity, and temperature) are considered. As a consequence, they sacrifice accuracy, in particular in the low collisionality regime \cite{pitzal2023}, which can be present in both the edge and SOL. An alternative (and more efficient) formulation of the full-$f$ GK model is based on a global and spectral velocity-space approach \cite{frei2020,frei2024}. Such an approach was recently implemented, for the first time, in diverted geometry in the full-$f$ GK turbulence code \texttt{GENE-X} \cite{frei2024}. Originally designed with an Eulerian velocity-space grid-based approach \cite{michels2021}, \texttt{GENE-X} now supports a spectral formulation in velocity-space. Equipped with this new velocity-space numerical discretization, it has been demonstrated in \cite{frei2024} that the spectral simulations achieved comparable outboard midplane (OMP) profiles (density, temperature, and radial electric field) to the grid-based ones in the TCV-X21 L-mode scenario \cite{oliveira2022,ulbl2023}, dominated by trapped electron mode (TEM) turbulence, with only a few spectral coefficients to describe the velocity space and, thus, enabling an unprecedented reduction in the computational cost. Despite these promising results, further evaluation of turbulence properties, transport levels, and providing a clear physical interpretation of the results remains to be addressed, which is the primary objective of this work.

Therefore, we extend the analysis and comparisons presented in \cite{frei2024} and provide a better physical understanding of the results of spectral simulations in the present study. We first examine the equilibrium (long-wavelength) gradients and perform a detailed analysis of the radial force balance, which has not been investigated in previous studies \cite{ulbl2023}. We demonstrate that the radial force balance is accurately satisfied in the \texttt{GENE-X} simulations, ensuring balance between radial electric fields, pressure gradients, and plasma flows (toroidal and poloidal). This is critical in a full-$f$ framework \cite{dif2009,zholobenko2021} where the slowly evolving and mean (long-wavelength) electric field interacts with rapid and short-wavelength fluctuations. Then, we explore how spectral resolution impacts turbulence and transport focusing in the edge region. To this end, we investigate the accuracy of the turbulence properties and associated transport predicted by spectral simulations by comparing them against the high-fidelity grid-based results from \cite{ulbl2023}. This prior study highlighted the importance of TEMs in edge temperature profiles and power balance. In this case, TEM-driven turbulence was identified through Fourier analysis of fluctuations and fluxes, revealing mode propagation, frequency spectra, consistent with TEM characteristics. We confirm that spectral simulations accurately capture all these TEM characteristics. These results align well with grid-based outcomes, with minimal dependence on spectral resolution. The power crossing the separatrix predicted by the spectral approach is shown to agree within $10\%$ of grid-based results and experimental measurements. Moreover, the SOL falloff length of the divertor heat flux is qualitatively reproduced. Additionally, we confirm that the electron-dominated heat flux in this scenario is entirely turbulent, with negligible contributions from the turbulence-enhanced diamagnetic (neoclassical) fluxes. Finally, we compare these results to Braginskii fluid simulations using the \texttt{GRILLIX} code \cite{stegmeir2019,zholobenko2021}, which, unlike GK simulations found to be in the TEM regime, are dominated by ITG mode-driven turbulence. 

The paper is organized as follows. In \sref{sec:edgeandsolturbulentmodel}, we present the edge and SOL GK turbulence model and describe both the grid-based \cite{michels2021} and spectral \cite{frei2024} approaches implemented in \texttt{GENE-X}. The numerical setup is outlined in \sref{sec:modellingsetup}. \Sref{sec:ompprofile} analyzes OMP gradients and the radial force balance. Edge turbulence properties and their comparison with grid-based results are discussed in \sref{sec:turbulencecharacteristics}, while \sref{sec:turbulenttranapsortanalysis} examines turbulent flux spectra, phase shifts, and power balance. A qualitative comparison with Braginskii fluid simulations is presented in \sref{sec:comparisonbraginskii}, and conclusions are given in \sref{sec:conclusions}.

\section{Edge and SOL Turbulence GK Model} \label{sec:edgeandsolturbulentmodel}

We begin by introducing the GK full-$f$ model for edge and SOL turbulence implemented in the \genex code. This model is formulated in two distinct approaches: the grid-based method, described in \sref{sec:gridgyrokineticmodel}, and the spectral method, detailed in \sref{sec:spectralgyrokineticmodel}. This section provides an overview of the key aspects of both approaches, while further details can be found in \cite{michels2021} and \cite{frei2024} for the grid-based and spectral formulations, respectively.

\subsection{The full-$f$ Grid-based Approach} \label{sec:gridgyrokineticmodel}

The full-$f$ GK model evolves the full gyroaveraged part of the gyrocenter distribution function, denoted by $ f_\alpha = f_\alpha(\bm{R}, v_\parallel, \mu,t)$, for particle species $\alpha$ of mass $m_\alpha$ and charge $q_\alpha$, in the gyrocenter phase-space described by the coordinates $(\bm{R}, v_\parallel, \mu, \theta)$. Here, $\bm{R}$ is the gyrocenter position, $v_\parallel = \bm{b} \cdot \bm{v}$ is the velocity parallel to the equilibrium magnetic field, and $\mu = m_\alpha v_\perp^2 / (2B)$ is the magnetic moment. The unit vector $\bm{b} = \bm{B} / B$ represents the direction of the equilibrium  magnetic field $\bm{B}$, while $v_\perp = \left|\bm{v} - \bm{b} v_\parallel\right|$ is the velocity component perpendicular to $\bm{B}$.

The evolution of $f_\alpha$ is obtained using the long-wavelength electromagnetic and collisional GK Vlasov equation given by \cite{brizard2007}

\begin{eqnarray} \label{eq:vlasov}
& \frac{\partial \left(  B_{\|}^*  f_\alpha  \right)}{\partial t}  + \nabla \cdot \left( B_{\|}^* \dot{\bm R} f_\alpha \right) +  \frac{\partial}{\partial v_{\|}}\left(   B_{\|}^* \dot v_\parallel f_\alpha \right)  \nonumber \\
& 
= \sum_{\beta} B_{\|}^* C_{\alpha \beta}(f_\alpha, f_\beta),
\end{eqnarray}
where the equations of motion are given by  

\begin{eqnarray}
\dot{\bm R} & = v_{\|}  \frac{\bm{B}^*}{B_{\|}^*}  + \frac{c}{q_\alpha B_{\|}^*} \bm{b} \times\left(\mu  \nabla  B + q_\alpha  \nabla \psi_{1 \alpha}\right), \label{eq:dotR} \\
\dot v_\parallel & =  - \frac{\bm{B}^*}{m_\alpha B_{\|}^*} \cdot\left(\mu  \nabla  B+q_\alpha \nabla \psi_{1 \alpha}\right)  - \frac{q_\alpha}{m_\alpha c} \frac{\partial A_{1 \parallel} }{\partial t} \label{eq:dotvpar},
\end{eqnarray}
with $\bm{B}^* =  \bm{B}_{T} + m_\alpha v_{\|} c \nabla \times \bm{b} / q_\alpha$. Here, $\bm B_{T} = \bm B  + \delta \bm B_\perp $ is the sum of the equilibrium magnetic field, $\bm B$, and the small amplitude fluctuating perpendicular magnetic field, $\delta \bm B_\perp \simeq \nabla A_{1 \|} \times \bm{b}$, with $A_{1 \|}$ the parallel component of the fluctuating magnetic vector potential. The gyrocenter Jacobian, $B_\parallel^* / m_\alpha$, is proportional to $B_{\parallel}^* = \bm b \cdot \bm B^* = B + m_\alpha v_{\|} c \bm b \cdot \nabla \times \bm{b}  / q_\alpha$. In \eref{eq:vlasov}, the generalized potential $\psi_{1 \alpha} = \phi_1 -  m_\alpha c^2 / (2 q _\alpha  B^2) \left|  \nabla_\perp \phi_1 \right|^2$ is the sum of the electrostatic potential $\phi_1$ and a second order correction term that ensures energy consistency \cite{scott2010}. Here, the perpendicular gradient is defined by $\nabla_\perp = (\mathbf{I} - \bm b \bm b ) \cdot \nabla$ (with $\mathbf{I}$ being the identity matrix). Finally, the effects of elastic Coulomb collisions between species $\alpha$ and $\beta$ are introduced through the collision operator $C_{\alpha \beta}$ on the right hand-side of \eref{eq:vlasov}. In this work, $C_{\alpha \beta}$ is modeled by a long-wavelength and full-$f$ Lenard-Bernstein/Dougherty (LBD) collision operator \cite{dougherty1964,ulbl2022}. 

The electromagnetic fields, $\phi_1$ and $A_{1 \parallel}$, are obtained self-consistently from the GK Maxwell equations, that are the quasineutrality (QN) condition and Ampere’s law, given by

 \begin{eqnarray} \label{eqs:fields}
- \nabla   \cdot\left(\sum_\alpha \frac{m_\alpha c^2}{B^2} \int d W f_\alpha \nabla_{\perp} \phi_1\right)  =\sum_\alpha q_\alpha \int d W f_\alpha ,  \label{eq:poisson}\\
-\Delta_{\perp} A_{1 \|}  = 4 \pi \sum_\alpha \frac{q_\alpha}{c} \int d W f_\alpha v_{\|}  \label{eq:ampere}.
 \end{eqnarray} 
On the other hand, the induction part of the parallel electric field, namely $\partial_t A_{1 \parallel}$, is treated as an independent variable and obtained from the Ohm's law, 

 \begin{eqnarray} \label{eq:ohmslaw}
& -\left(\Delta_{\perp}+ 4 \pi \sum_\alpha \frac{q_\alpha^2}{m_\alpha c^2} \int d W \frac{\partial f_\alpha}{\partial v_{\|}}  \right)  \frac{\partial A_{1 \parallel} }{\partial t}  \nonumber \\ 
& = 4 \pi \sum_\alpha \frac{q_\alpha}{c} \int d W  \left(\frac{\partial f_\alpha}{\partial t}\right)^{\star} v_{\|}.
\end{eqnarray}
In \eref{eq:ohmslaw}, the superscript $\star$ indicates the part of $\partial_t f_\alpha $ which is independent of $\partial_t A_{1 \parallel}$, i.e. $(\partial_t f_\alpha)^\star  = \partial_t f_\alpha  - q_\alpha / (m_\alpha c) \partial_{v_{\|}} f_\alpha \partial_t A_{1 \parallel} $. 

The set of equations \eref{eq:vlasov}, \eref{eqs:fields}, and \eref{eq:ohmslaw} constitutes a closed system of partial differential equations for the full-$f$ distribution function $f_\alpha$ and the perturbed electromagnetic fields $\phi_1$ and $A_{\parallel 1}$ (as well as $\partial_t A_{\parallel 1}$). These equations were originally implemented in \genex using a grid-based approach in velocity-space \cite{michels2021,michels2022}. In this approach, the velocity-space $(v_\parallel, \mu)$ is discretized on a fixed regular grid constituted by $(N_{v_\parallel}, N_\mu)$ grid points in the $v_\parallel$ and $\mu$ directions, respectively. Velocity-space derivatives and integrals are then approximated using finite difference schemes and quadrature formulas. In the grid-based approach, the velocity-space grid spacing decreases linearly with the number of grid points $N_{v_\parallel}$ and $N_\mu$. Therefore, any fine and localized structures in velocity-space can be resolved by increasing $N_{v_\parallel}$ and $N_\mu$. However, a large number of grid points is often required to properly resolved the entire velocity-space. This is particularly pronounced in the presence of large temperature difference in the edge. It is found that typically that $(N_{v_\parallel} , N_\mu ) \sim (80, 20)$ are required to achieve well resolved simulations in medium-sized devices (e.g., TCV \cite{ulbl2023} and AUG \cite{michels2022}). In this work, we consider the grid-based approach as reference in order to compare the spectral approach, which is introduced below.

\subsection{The full-$f$ Spectral Approach} \label{sec:spectralgyrokineticmodel}

Contrary to the grid-based approach where the distribution function is solved on a fixed velocity-space grid, the spectral formulation, introduced recently in \cite{frei2024}, expands $f_\alpha$ onto a set of orthogonal and scaled Hermite and Laguerre polynomials. More precisely, the distribution function, $f_\alpha$, is approximated by \cite{frei2020}

\begin{equation} \label{eq:faspectral}
f_\alpha \simeq \sum_{p =0}^{N_{v_\parallel}-1} \sum_{j =0}^{N_\mu-1} \mathcal{N}_\alpha^{pj} \hat H_p(\hat v_{\parallel \alpha}) L_j(\hat \mu_\alpha) F_{\mathcal{M} \alpha},
\end{equation}
where $\hat H_p(\hat v_{\parallel \alpha}) = H_p(\hat v_{\parallel \alpha})  / \sqrt{2^p p !} $ and $L_j(\hat \mu_\alpha)$ are the Hermite and Laguerre polynomials \cite{gradshteyn2014} with the scaled velocity-space coordinates $\hat v_{\parallel \alpha} = v_\parallel \sqrt{2 \tau_\alpha / m_\alpha}$ and $\hat \mu_\alpha = \mu B  / \tau_\alpha$ as arugments. Here, $\tau_\alpha$ a constant reference temperature introduced to adjust the velocity-space basis (see Appendix C of \cite{frei2024} for more details). In \eref{eq:faspectral}, the spectral coefficients, $\mathcal{N}_\alpha^{pj}$, are defined as Hermite-Laguerre weighted moments of $f_\alpha$ and can be related to fluid quantities \cite{frei2024}. We note that, in \eref{eq:faspectral}, the sums are truncated for numerical reasons up to $p = N_{v_\parallel} -1 $ and $j = N_{\mu} -1 $ while \eref{eq:faspectral} becomes an equality if $N_{v_\parallel}, N_\mu \to \infty$. As a consequence, $N_{v_\parallel} \times N_\mu$ spectral coefficients are evolved per species. 

In the spectral approach, the self-consistent evolution of $f_\alpha$ is obtained from the evolution of the spectral coefficients $\mathcal{N}_\alpha^{pj}$, which are related to Hermite-Laguerre weighted velocity-space moments of $f_a$. By projecting \eref{eq:vlasov} onto the scaled Hermite-Laguerre polynomial basis, we derive \cite{frei2024}

\begin{equation} \label{eq:spectralvlasov}
    \frac{\partial}{\partial t } \mathcal{N}^{pj}_\alpha  + \nabla  \cdot \bm \Gamma_\alpha^{pj} + \mathcal{F}_{\alpha \ell k}^{pj}\mathcal{N}^{\ell k}_\alpha + \mathcal{D}_{\alpha \ell k}^{pj}\mathcal{N}^{\ell k}_\alpha \frac{\partial A_{1 \parallel} }{\partial t} = \sum_{\beta} \mathcal{C}_{\alpha \beta}^{pj},
\end{equation}
where the generalized flux $\bm \Gamma_\alpha^{pj}$ and matrices, $\mathcal{F}_{\alpha \ell k}^{pj}$ and $\mathcal{D}_{\alpha \ell k}^{pj}$, depend non-linearly on $\mathcal{N}^{pj}_\alpha$ and on the electromagnetic fields. On the right hand-side of \eref{eq:spectralvlasov}, collisional effects, $\mathcal{C}_{\alpha \beta}^{pj}$, are modeled using a spectral full-$f$ Lernard-Bernstein Daugherty collision operator \cite{ulbl2022,frei2024}. The analytical expressions of $\Gamma_\alpha^{pj}$, $\mathcal{F}_{\alpha \ell k}^{pj}$, $\mathcal{D}_{\alpha \ell k}^{pj}$, and $\mathcal{C}_{\alpha \beta}^{pj}$ can be found in \cite{frei2024}. 

\Eref{eq:spectralvlasov} is coupled to the spectral GK Maxwell equations which are deduced from \eref{eqs:fields} and \eref{eq:ohmslaw} using \eref{eq:faspectral} to evaluate analytically the velocity-space integrals. This leads to the spectral QN condition, Ampere's and Ohm's law, given by 

\begin{eqnarray} 
 -\boldsymbol{\nabla} \cdot\left( \sum_\alpha   \frac{  m_\alpha  c^2 \mathcal{N}_\alpha^{00}}{B^2}   \nabla_{\perp}  \phi_1\right)=\sum_\alpha q_\alpha \mathcal{N}_\alpha^{00}   \label{eq:poissonvspec} ,\\
 -  \Delta_{\perp}  A_{1 \|}  =   4 \pi  \sum_\alpha  \frac{q_\alpha}{c}   \sqrt{ \frac{\tau_{\alpha}}{  m_\alpha}}    \mathcal{N}_\alpha^{10}  \label{eq:amperevspec} , \\
   - \left(   \frac{\Delta_{\perp} }{4 \pi} -   \sum_\alpha \frac{q_\alpha^2 \mathcal{N}_\alpha^{00}}{c^2 m_\alpha}   \right)\frac{\partial A_{1 \parallel} }{\partial t} =    \sum_\alpha \frac{q_\alpha}{c}   \sqrt{ \frac{\tau_{\alpha}}{  m_\alpha}}  \left( \frac{\partial \mathcal{N}_\alpha^{10}}{\partial t}\right)^{\star} \label{eq:ohmvspec},
\end{eqnarray}
respectively. In \eref{eq:ohmvspec}, $(\partial_t \mathcal{N}_\alpha^{10})^{\star}$ denotes the part of $\partial_t \mathcal{N}_\alpha^{10}$ which is independent of $\partial_t A_{1 \parallel}$. We remark that the self-consistent density is retained in the polarization term contained in the QN condition appearing on the left hand-side of \eref{eq:poissonvspec}. This is in contrast to previous TCV-X21 investigations using \texttt{GENE-X} which uses a linearized polarization \cite{ulbl2023}.

From the spectral expansion  \eref{eq:faspectral}, the particle charge density and parallel current can be related to the spectral coefficients $\mathcal{N}_\alpha^{00}$ and $\mathcal{N}_\alpha^{10}$ via their definitions such that $n_\alpha = \int d W  f_\alpha  = \mathcal{N}_\alpha^{00}$ and $n_\alpha u_{\parallel \alpha }  =  \int d W v_{\| } f_\alpha  =  \sqrt{ \tau_{\alpha} / m_\alpha }   \mathcal{N}_\alpha^{10}$. Similarly, the total temperature $T_\alpha = ( T_{\parallel \alpha}  + 2 T_{\perp \alpha})/3 $ can be computed using $n_\alpha  T_{\parallel \alpha}  =   \tau_\alpha  \left(  \sqrt{2} \mathcal{N}_\alpha^{20} +  n_\alpha \right)  -  n_\alpha  m_\alpha  u_{\parallel \alpha } ^2$ and $n_\alpha T_{\perp \alpha} =  \tau_{\alpha}   \left(n_\alpha - \mathcal{N}_\alpha^{01} \right)$. Similar relations can be obtained for the heat fluxes and are given in \cite{frei2024}.

In order to solve numerically the spectral GK Vlasov equation given in \eref{eq:spectralvlasov}, a closure by truncation is applied. This is done by setting the spectral coefficients $\mathcal{N}_\alpha^{pj}$ to zero for $p \ge N_{v_\parallel}$ and $j \ge N_{\mu}$. Additionally, a constant diagonal damping coefficient is introduced on the right-hand side of \eref{eq:spectralvlasov} to mitigate the effects of finite spectral resolution \cite{frei2023}. The combination of this diagonal damping term and truncation closure has proven to be sufficiently numerically robust \cite{frei2024}.  

In \cite{frei2024}, it was shown that a small number of spectral coefficients, in that case $(N_{v_\parallel}, N_\mu) \sim (6,4)$ was sufficient to achieve a good agreement in the OMP profiles with respect to the grid-based simulations \cite{ulbl2023}. This number is at least an order of magnitude smaller than the number of velocity-space grid points required in the grid-based approach. Depending on the chosen method (grid-based or spectral), we refer to the velocity-space resolution as $(N_{v_\parallel}, N_\mu)$. In the following, we compare the turbulence predictions obtained using the spectral approach, presented in this section, with different $(N_{v_\parallel}, N_\mu)$ values against those from the grid-based formulation detailed in \sref{sec:gridgyrokineticmodel}.

\section{Modelling Setup of TCV-X21} \label{sec:modellingsetup}

All simulations in this study are based on the TCV-X21 experimental scenario which is a well-established L-mode deuterium lower-single-null configuration for turbulence code validation \cite{oliveira2022,ulbl2023,frei2024}. The experimental details of this scenario are thoroughly documented in \cite{oliveira2022}. While the corresponding numerical setup for the \texttt{GENE-X} simulations are reported in \cite{ulbl2023,frei2024}, we recall here only the relevant details for the present study. We remark that all the data of the grid-based simulations are available in \cite{tcvx21zenodo}.

In the grid-based simulations reported in \cite{ulbl2023}, a velocity-space resolution of $(N_{v_\parallel}, N_\mu) = (80, 60)$ was employed, with an equidistant grid in $\mu$. While no velocity-space resolution scan was conducted (in a similar fashion as in \cite{frei2024} for spectral simulations), numerical tests suggested that $N_\mu$ could be reduced to approximately $20$, without significantly compromising accuracy \cite{ulbl2023phd}. Given that our comparison builds on the analysis on the turbulence analysis of \cite{ulbl2023}, we treat the $(80, 60)$ grid-based simulations as the highest-fidelity reference case in the present study. In contrast, the spectral simulations in \cite{frei2024} utilized significantly fewer spectral coefficients. More precisely, we consider the results from the spectral resolutions of $(N_{\mu}, N_{v_\parallel}) = (4,2)$, $(8,4)$, and $(16,8)$.

Both spectral and grid-based approaches use numerically motivated boundary conditions. More precisely, Dirichlet boundary conditions are applied at the inner and outer radial real space domain boundaries where the distribution functions are set to local Maxwellians whose density and temperature match the experimental profiles. This choice of boundary conditions imposes that odd spectral coefficients, associated with parallel particle and heat fluxes, vanish at the boundaries (see Appendix in \cite{frei2024}). Homogeneous Dirichlet conditions are also imposed on the electromagnetic fields $\phi_1$ and $A_{\parallel 1}$. A buffer region with second order diffusion is introduced near the boundaries to enhance numerical stability.

The reference parameters for all simulations are chosen to be $T_{\mathrm{ref}} = 20$~eV, $n_{\mathrm{ref}} = 10^{19}$~m$^{-3}$, $B_{\mathrm{ref}} = 0.929$~T, and $L_{\mathrm{ref}} = R_0 = 0.906$~m, with $R_0$ the major radius of TCV. The simulations consider a deuterium plasma with $m_{\mathrm{i}} = 2 m_{\mathrm{p}}$ ($m_{\mathrm{p}}$ is the proton mass) and a realistic electron mass $m_{\mathrm{e}} \simeq 1/1830 m_{\mathrm{p}}$. The computational grid uses $32$ poloidal planes in $\varphi$ with a uniform spacing of $1.23$~mm in $RZ$, resulting in $200657$ grid points per poloidal plane. All simulations, both spectral and grid-based, reach a quasi-steady state after approximately $0.5$ ms. Finally, the exact computational reduction achieved by the spectral approach in this case are reported in \cite{frei2024}.

\section{Outboard Midplane Gradients and Electric Field Analysis} \label{sec:ompprofile}

We first examine the gradients of the OMP profiles in \sref{sec:omp}. While it has been demonstrated that the spectral approach accurately reproduces the OMP profiles \cite{frei2024}, even minor discrepancies in their gradients could lead to significant differences in the turbulence regime, which is the primary focus of this work. The consistency of \texttt{GENE-X} in fulfilling the radial force balance in the edge region - a critical aspect of full-$f$ GK modeling - is also investigated in \sref{sec:forcebalance}. Finally, the composition of the radial electric field is analyzed in detail (\sref{sec:compoedge}) and in the SOL (\sref{sec:composol}) using analytical predictions that we derive from the GK equation.

\subsection{OMP Gradients Analysis} \label{sec:omp}

\begin{figure}
    \centering
    \includegraphics[scale=0.5]{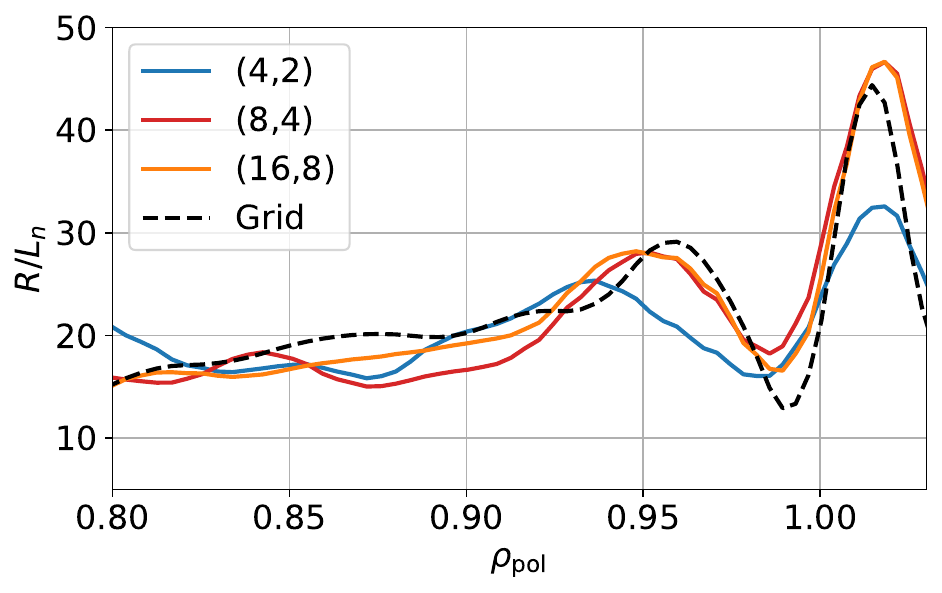}
    \includegraphics[scale=0.5]{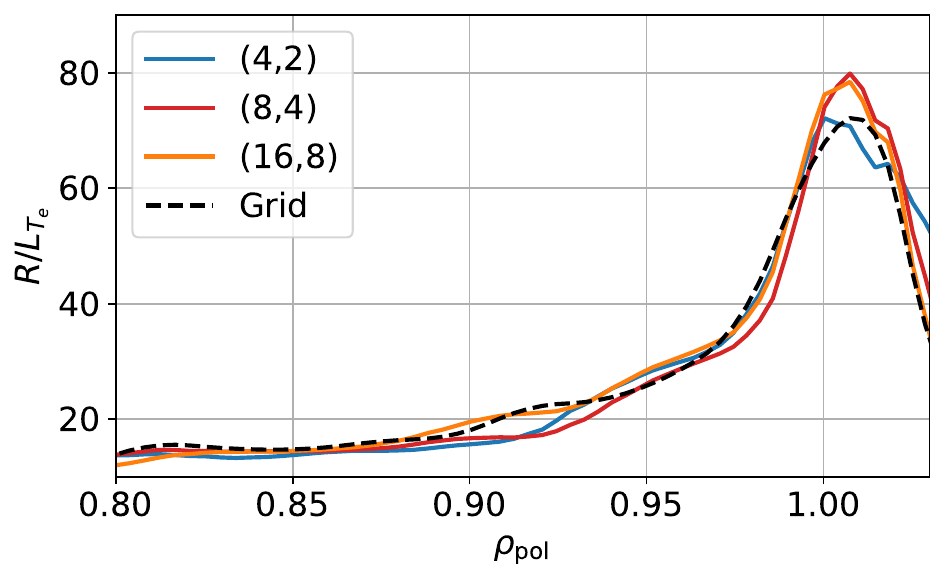}
    \includegraphics[scale=0.5]{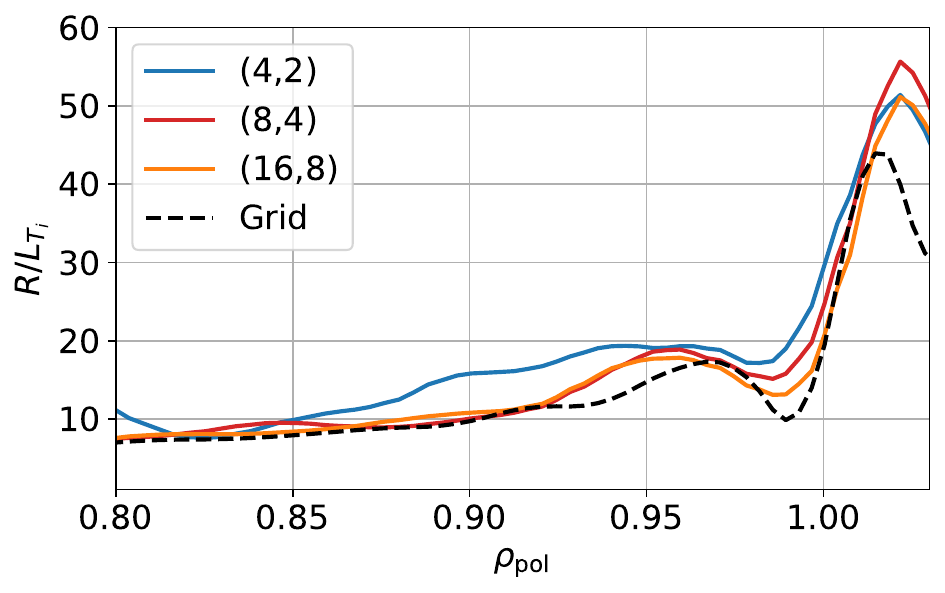}
    \caption{OMP normalized gradients of density $n$ (top), electron temperature $T_e$ (middle), and ion temperature $T_i$ (bottom) shown as a function of the normalized flux-surface label $\rho_\mathrm{pol}$. The colored lines are the results from the spectral simulations \cite{frei2024} and the dashed black are the gradients obtained from the grid-based simulation \cite{ulbl2023}.}
    \label{fig:ompgradients}
\end{figure}

From the OMP profiles of the spectral simulations reported in \cite{frei2024}, the associated normalized density and temperature gradients, defined as $L_{n}/R = - (\nabla r \cdot \nabla \ln n)^{-1} / R$ and $L_{T}/R = - (\nabla r \cdot \nabla \ln T)^{-1} / R$ (where $\nabla r$ indicates the radial direction and $R$ is the major radius of TCV), are computed and presented in \fref{fig:ompgradients}. The normalized gradients from both spectral and grid-based simulations are compared. 

First, we observe that, similarly to the OMP profiles, the spectral approach demonstrates good overall agreement in reproducing the normalized gradients. In particular, the positions and magnitudes of the gradient peaks are accurately captured, in particular for the electron temperature gradient in the edge region. While minor differences are observed between the $(16,8)$ and $(8,4)$ cases, small deviations in amplitude emerge in the $(4,2)$ case, particularly for the density and ion temperature gradients. These findings align with the OMP profiles reported in \cite{frei2024}, where larger deviations were noted in $n$ and $T_\mathrm{i}$, while the $T_\mathrm{e}$ profiles remained nearly identical across all spectral simulations, a trend also reflected in the electron temperature gradients. 

Second, all gradients increase toward the separatrix and remain relatively constant further into the edge region ($\rho_\mathrm{pol} = 0.90$). Beyond $\rho_\mathrm{pol} = 0.90$, the electron temperature gradient exceeds the ion temperature gradient, with $R / L_{T_e} \simeq 20$ compared to $R / L_{T_i} \simeq 10$. This suggests that turbulence in this region is likely driven by density and/or electron temperature gradients, as ion temperature gradient (ITG) modes are expected to be linearly stable in the edge, given $\eta_i = L_{N} / L_{T_i} \lesssim 0.5$ \cite{frei2022}. 

Overall, the analysis of \fref{fig:ompgradients} indicates that possible deviations in the edge turbulence properties (and transport) between the spectral and grid-based simulations are unlikely to be related to deviations in the equilibrium gradients.

\subsection{Radial Force Balance Analysis} \label{sec:forcebalance}

We now investigate the consistency of radial force balance in the edge region. This is crucial for full-$f$ GK modeling, where both the long-wavelength, slowly evolving radial electric field and plasma flows (toroidal and poloidal) interact with small perpendicular wavelength and rapid turbulent fluctuations.

\begin{figure}
    \centering
    \includegraphics[scale=0.5]{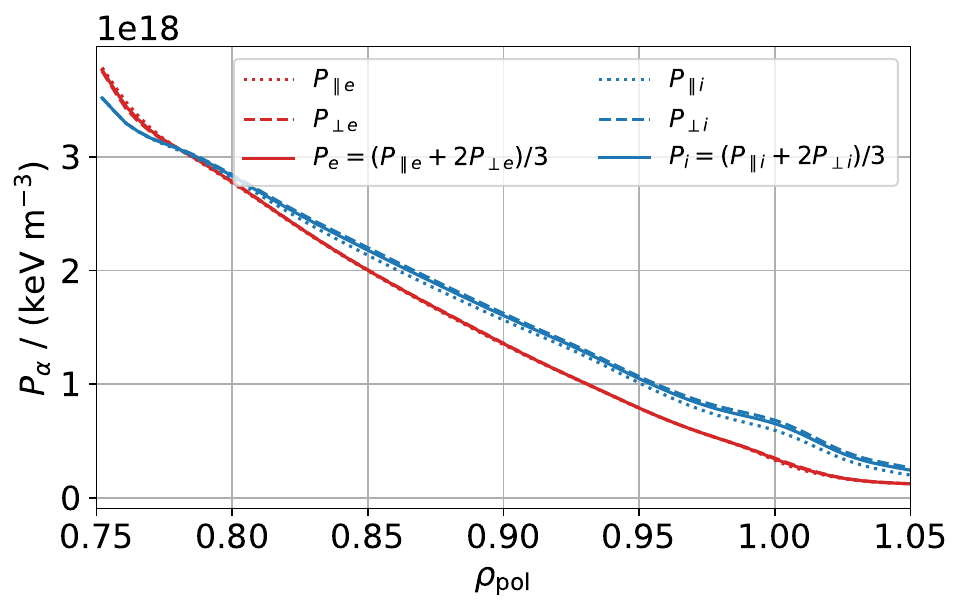}
     \caption{OMP parallel ($P_{\parallel \alpha}$), perpendicular ($P_{\perp \alpha}$) and total ($P_{ \alpha}$) pressure profiles for the electrons (red) and ions (blue). The profiles are averaged over the toroidal direction and $0.1$~ms at quasi-steady state obtained from the $(16,8)$ spectral simulation. Similar profiles are obtained with $(8,4)$ and $(4,2)$. }
    \label{fig:pressureomp}
\end{figure}

In the closed field line region, the radial electric field, $E_r = \nabla r \cdot \bm{E} = - \partial_r \phi_1$, must be balanced by the toroidal and poloidal flows as well as by the radial pressure gradient. In the absence of inertia and friction forces, the radial force balance, deduced from the radial projection of the ion momentum equation, reads \cite{hazeltine2013}

\begin{equation} \label{eq:GKforcebalance}
E_r = \frac{1}{ q_i n_i} \left( \nabla \cdot \bm{\Pi}_i \right) \cdot \nabla r -   U_{i \theta} B_\phi +  U_{i \phi} B_\theta, 
\end{equation}
where the anisotropic ion pressure tensor $\bm{\Pi}_i$ is given by

\begin{equation} \label{eq:pressuretensor}
\bm \Pi_i = P_{\parallel i} \bm b \bm b +  P_{\perp i} \left( \mathbf{I} - \bm b \bm b\right).
\end{equation} 
In \eref{eq:GKforcebalance}, $U_{i \theta} = \bm e_{\theta} \cdot \bm U_i$ and $U_{i \phi}= \bm e_{\phi} \cdot \bm U_i$ are the poloidal and toroidal components of the plasma flow $\bm U_i$. Similarly, $B_{\phi}$ and $B_{\phi}$ are the toroidal and poloidal components of the magnetic field $\bm B$, with $B_{\theta} \ll B_\phi \sim B$. From \eref{eq:GKforcebalance}, it follows that $E_r$ must balance the radial ion pressure gradient, if the poloidal and toroidal plasma flows are small. Otherwise, finite poloidal rotation, which is independent of $E_r$ \cite{kim1991}, establishes. We remark that \eref{eq:GKforcebalance} encompasses both neoclassical and turbulence-induced contributions to the radial electric field and plasma flow. Also, we note that the radial force balance given in \eref{eq:GKforcebalance} differs from the more classical force balance equation since the former allows for anisotropic pressure. However, due to the collisional energy exchange rate, a small difference (of the order or less than $10\%$) is observed in the present case, as illustrated in Figure \ref{fig:pressureomp}. Therefore, by considering $P_{\parallel i} \simeq P_{\perp i} \simeq P_i$, the pressure tensor becomes $\bm \Pi_i = \mathbf{I} P_i$, such that $\left(\nabla \cdot \bm \Pi_i \right) \cdot \nabla r \simeq \partial_r P_i$. We note that the effect of pressure anisotropy may be non-negligible in high-performance scenarios, such that the correct expression of $\bm \Pi_i$ in \eref{eq:GKforcebalance} must be used.

To investigate the radial force balance, we measure the contribution of the poloidal flow, proportional to $U_{i \theta} B_{\phi } $, instead of $E_r$ which is studied in more detail in the next section. Assuming isotropic pressure (see \fref{fig:pressureomp}), an expression for $U_{i \theta}$ can be deduced from the force balance equation given in \eref{eq:GKforcebalance}, such that

\begin{equation} \label{eq:poloidalflow}
 U_{i \theta} B_\phi = \frac{1}{n_e q_i} \partial_r P_i  - E_r +  U_{i \phi} B_\theta.
\end{equation}
\Eref{eq:poloidalflow} yields a constraint on the poloidal flow which self-consistently establishes in the simulations to fulfill the radial force balance given in \eref{eq:GKforcebalance}. In addition, the poloidal flow, expressed in \eref{eq:poloidalflow}, contains contributions from both the long-wavelength, slow neoclassical components and the shorter-wavelength, rapid turbulence part. 

To verify \eref{eq:poloidalflow}, we compute the different terms on the right hand-side separately. More precisely, we calculate the toroidal flow by assuming that it is mainly driven by the parallel ion velocity, such that $U_{\phi i } \simeq u_{\parallel i}$. Then, the radial electric field and the radial ion pressure gradient are obtained directly from $\phi_1$, such that $E_r = - \partial_r \phi_1$, and from the total ion pressure $P_i = n_i (T_{\parallel i} + 2 T_{\perp i}) / 3 $. On the other hand, the poloidal flow appearing on the left hand-side of \eref{eq:poloidalflow} is computed self-consistently from the simulation by taking the poloidal projection of the total ion plasma flow, which is defined as the sum of the gyrocenter drift (composed by the $\bm E \times \bm B$ and magnetic drifts) and a classical magnetization. More precisely, the poloidal projection of the plasma flow, $U_{i \theta } = \bm U_i \cdot \bm e_\theta$, is expressed as \cite{hazeltine2013}

\begin{equation} \label{eq:poloidaldrifts}
    U_{i \theta } =  V_{i \theta}  + \bm e_\theta \cdot \left(\frac{c}{q_i n_i} \nabla \times \bm m_i \right).
\end{equation}
Here, $V_{i \theta}  = \bm e_\theta \cdot \bm V_{i}$ is the poloidal component of the total gyrocenter drift, $\bm V_{i} = \int d W \dot \bm{ R} f_i / n_i$ with $\dot \bm R$ given in \eref{eq:dotR}. From the equation of motion $\dot \bm{ R}$ in \eref{eq:dotR}, the total gyrocenter drift, $\bm V_{i}$, can be written as follows
 
\begin{eqnarray} \label{eq:gcdrift}
\bm V_{i} & = u_{\parallel i}\bm b +  \frac{c \bm E \times \bm B}{B^2} \nonumber \\ 
& + \frac{c  T_{\parallel i}}{q_i B} \nabla \times \bm b   + \frac{c T_{\perp i}}{q_\alpha B^2 }   \bm b \times \nabla B.
\end{eqnarray}
Here, the contributions from the flutter drift as well as from the second-order Hamiltonian are ignored since they both yield only small corrections to $\bm V_{i}$. We use \eref{eq:gcdrift} inserted into \eref{eq:poloidaldrifts} to evaluate the self-consistent poloidal flow rotation $  U_{i \theta } $ in the spectral simulations. 

The last term in \eref{eq:poloidaldrifts}, proportional to the rotational of $\bm m_i  = - P_{\perp i} \bm b / B$, is the classical magnetization correction which arises due the gyrophase dependent part of the gyrocenter distribution function \cite{hazeltine2013}. We remark that, since the classical magnetization contribution is divergence free, the gyrocenter drift $\bm V_{i}$ is equal to the plasma flow only under the divergence operator, i.e. $\nabla \cdot \bm U_i = \nabla \cdot \bm V_i$, such that they both yield the same transport across a flux-surface. However, the classical magnetization correction to the gyrocenter drift is important when considering the poloidal flow as we show below to establish the radial force balance.

\begin{figure}
    \centering
    \includegraphics[scale=0.5]{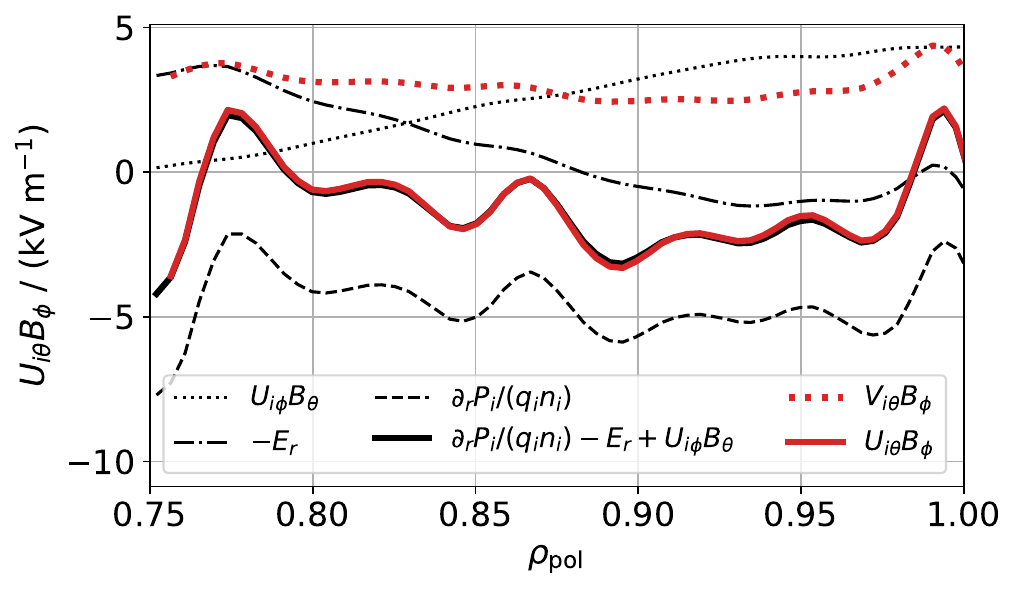}
     \caption{Poloidal flow contribution to the force balance at the OMP position estimated from (solid black) the force balance given in \eref{eq:poloidalflow} and (red line) self-consistently computed from the spectral $(16,8)$ simulation using \eref{eq:poloidaldrifts}. The radial force balance is accurately satisfied. The quantities are evaluated on a single poloidal plane and average over $0.1$~ms in steady-state. The black non-solid line indicate the different contributions in \eref{eq:poloidalflow} while the red dotted line indicates $V_{i \theta}$ given in \eref{eq:gcdrift}. Negative (positive) values indicate the ion (electron) diamagnetic direction.}
    \label{fig:poloidalflow}
\end{figure}

In \fref{fig:poloidalflow}, we show $U_{i \theta} B_\phi$ estimated using \eref{eq:poloidalflow} and self-consistently calculated using \eref{eq:poloidaldrifts} in the spectral $(16,8)$ simulation at the OMP position. The contributions to the poloidal flow are also displayed to emphasize the dominant driving mechanisms as well as the contribution of the poloidal projection gyrocenter drift $V_{i \theta}$ to $U_{i \theta}$. The results are averaged over $0.1$~ms on a single poloidal plane under steady-state conditions. The results confirm that the force balance is well satisfied. We note that, while the results predicted in \fref{fig:poloidalflow} are time-averaged, the force balance remains consistent at all time in the simulations. The small deviations observed may arise from the toroidal flow approximation and numerical noise. Finally, we notice the significant departure of the poloidal projection of $\bm V_i$, which closely follows $- E_r$ (shown by the dashed-dotted black line), demonstrates that the inclusion of the classical magnetization (the last term in \eref{eq:poloidaldrifts}) is critical to accurately predict the poloidal plasma flow, as imposed by the radial force balance.

The force balance analysis presented here is robust and reproducible, as demonstrated by consistent results obtained from $(4,2)$ and $(8,4)$ spectral simulations \cite{frei2024}. In addition, the findings reported here are equally applicable to the grid-based approach \cite{ulbl2023}. Therefore, this present analysis confirms that the full-$f$ simulation of \texttt{GENE-X} accurately captures the radial force balance, which governs the mean and long wavelength radial electric field in the edge region through its interplay with the equilibrium radial pressure gradient, as well as (neoclassical and turbulent) flows \cite{dif2009,zholobenko2021}.  

\subsection{Composition of $E_r$ in the Edge}\label{sec:compoedge}

Having established the radial force balance, we now focus on analyzing the composition of the radial electric field in the edge. To highlight the contribution of the poloidal flow $U_{i \theta}$ in the radial electric field, we can estimate of $E_r$ from \eref{eq:GKforcebalance} as follows

\begin{equation} \label{eq:eresimtate}
 E_r \simeq \frac{1}{q_i n_i} \partial_r P_i + U_{i \phi} B_\theta.
\end{equation}
Therefore, by construction, any deviations between $E_r$ and \eref{eq:eresimtate} are attributed to the presence of poloidal (turbulence-induced and/or neoclassical) flows. 

\begin{figure}
    \centering
    \includegraphics[scale=0.5]{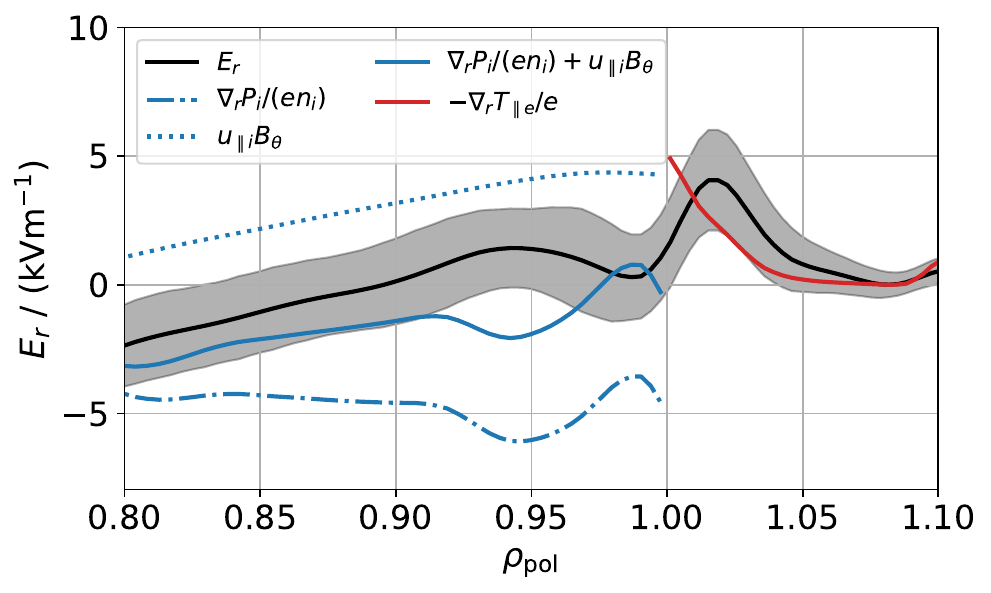}
     \caption{Composition of the radial electric field $E_r$ (solid black line) at the OMP from the spectral $(16,8)$ simulation. The solid blue line represent the estimate of $E_r$ from force balance \eref{eq:eresimtate}, with contributions from the total ion pressure gradient (dashed-dotted) and intrinsic toroidal rotation (dotted). The solid red line indicates the SOL estimate of $E_r$, obtained using \eref{eq:ersol}.}
    \label{fig:omper}
\end{figure}

\Fref{fig:omper} shows the OMP profile of the radial electric field $E_r$ obtained from the spectral $(16,8)$ simulation. The individual contributions of the ion pressure gradient and toroidal rotation are also shown. First, we observe that $E_r$ closely follows \eref{eq:eresimtate} within the range $0.8 \lesssim \rho_{\mathrm{pol}} \lesssim 0.9$, where both the toroidal rotation and ion pressure gradient have comparable contributions. In particular, the toroidal rotation plays a crucial role in determining $E_r$, yielding an upper shift of $E_r$ further inside the edge while the ion pressure gradient tends to decrease $E_r$. We remark that, this is consistent with the fact that, in L-mode plasmas, deviations from the leading-order relation, $E_r \sim \nabla_r P_i / (n_e e)$, are frequently observed in experiments \cite{plank2023}. While the $E_r$ profile approximately follows \eref{eq:eresimtate} further inside the edge ($\rho_{\mathrm{pol}} \lesssim 0.9$), large deviations from \eref{eq:eresimtate} occur near the separatrix ($\rho_{\mathrm{pol}} \gtrsim 0.9$) due to the presence of non-negligible poloidal flows, as shown in \fref{fig:poloidalflow}.

It is worth mentioning that the decrease of $E_r$ around $\rho_{\mathrm{pol}} \lesssim 0.85$ is due to the weakening of the toroidal rotation contribution imposed by the the inner Dirichlet boundary condition. This choice of boundary condition sets all flux variables (including $u_{\parallel i}$) to zero \cite{frei2024}, explaining the decrease of the toroidal rotation towards the inner boundary. Numerical tests suggest that using a Neumann boundary condition for $u_{\parallel i}$ at the edge has a minor effect on $E_r$ in the present case. However, proper inner boundary conditions for finite parallel flow may be more important in neutral beam injection heated plasmas and high-confinement regimes, where $u_{\parallel i}$ can be larger and enhance $E_r$ at the edge.

\subsection{Composition of $E_r$ in the SOL} \label{sec:composol}

Contrary to the closed field line region where it is the radial force balance that dictates $E_r$, it is the parallel electron dynamics that dominates in the SOL. Therefore, to obtain an analytical estimate for $E_r$ in the SOL, we derive a generalized Ohm's law from the GK Vlasov equation given in \eref{eq:vlasov}. More precisely, we consider the electron momentum equation by setting $(p,j) = (1,0)$ in \eref{eq:spectralvlasov}. In the low-beta limit and at the leading order in $m_i / m_e$, it yields

\begin{equation} \label{eq:pj10}
   P_{\Delta e}\nabla_{\parallel} \ln B + \nabla_\parallel P_{\parallel e} - e n_e \nabla_\parallel \phi_1 = \sum_{\beta} \sqrt{\tau_e m_e}\mathcal{C}^{10}_{e \beta},
\end{equation}
where $P_{\Delta e} = P_{\parallel e} - P_{\perp e}$ measures the electron pressure anisotropy. The evaluation of the friction force on the right hand-side of \eref{eq:pj10} is obtained from the explicit expressions of the LBD collision operator reported in \cite{frei2024}, yielding $\sqrt{\tau_e m_e}\mathcal{C}_{e \beta} = m_e n_e \nu_{ei}  \left( u_{\parallel i } - u_{\parallel e}\right)$. This enable us to write \eref{eq:pj10} as 
 
\begin{equation} \label{eq:generalizedohmslaw}
   \frac{P_{\Delta e}}{en_e} \nabla_{\parallel} \ln B +  \frac{1}{en_e} \nabla_\parallel P_{\parallel e} -  \nabla_\parallel \phi_1 =  \eta_{\parallel} j_{\parallel},
\end{equation}
introducing the parallel resistivity $\eta_{\parallel} = m_e \nu_{ei} / (e^2 n_e)$ and the parallel current $j_{\parallel} = e n_e (u_{\parallel i} - u_{\parallel e})$. 

Before analyzing the composition of the electric field in the SOL using \eref{eq:generalizedohmslaw}, it is worth noticing the difference between \eref{eq:generalizedohmslaw}, derived from the collisional GK Vlasov equation, and the generalized Ohm's law in Braginskii fluid model (see, e.g., \cite{zholobenko2021}). The first difference is the presence of the electron parallel and perpendicular pressures (in the first two terms), which are not present in the Braginskii model due to the isotropic pressure assumption. However, both in the SOL and in the edge, the electron pressure anisotropy is found to be weak such that $P_{\Delta e} \simeq 0$, as observed in \fref{fig:pressureomp}. A second difference is the absence in \eref{eq:generalizedohmslaw} of the thermal friction force proportional to $0.71 \nabla_\parallel T_e$,. This is a consequence of the collision operator model as the $\nabla_\parallel T_e$ term arises from the velocity dependence of the collision frequency, not retained in the LBD collision operator model. Another consequence of using the LBD collision operator is the difference in the prefactor ($0.51$ in the Braginskii model) contained in the parallel resistivity $\eta_\parallel$ (it is $1$ using the LBD collision operator).

To obtain an expression of the radial electric field in the SOL at the OMP, we integrate \eref{eq:generalizedohmslaw} along the magnetic field lines, from the divertor position $l_d$ to the OMP position $l_O$ (here, $l$ is the arc length parametrizing the magnetic field line). For simplicity, the pressure anisotropy and density gradient are neglected. We derive 

\begin{equation} \label{eq:phiomp}
  \phi_{1}(l_O)\simeq  \int_{l_{d}}^{l_{O}} d l \eta_{\parallel} j_{\parallel} + \frac{1}{e} \left(T_{\parallel e} (l_O) - T_{\parallel e}(l_d) \right),
\end{equation}
where we use that $\phi_{1}(l_d) = 0$ imposed by the boundary condition at the divertor plate. Evaluating the radial gradient of \eref{eq:phiomp}, while neglecting the $\eta_\parallel j_\parallel$ term, leads to an estimate of the radial electric field in the SOL, i.e.

\begin{equation} \label{eq:ersol}
  E_r \simeq - \frac{1}{e} \nabla_r  T_{ \parallel e} (\ell_O) 
\end{equation}
An interesting feature in \eref{eq:ersol} which appears from the GK treatment of the electron dynamics, is that, unlike the Braginskii model (see, e.g., \cite{zholobenko2021}), $E_r$ follows the radial gradient of the parallel electron temperature $T_{ \parallel e}$, rather than the total temperature $T_e$. However, as argued before, electron pressure anisotropy is found to be small such that $T_{\parallel e} \simeq T_e$ (see \fref{fig:pressureomp}). From \Eref{eq:ersol}, we deduce that $E_r \sim -  \nabla_r T_{\parallel e} (\ell_O) / e > 0$ is positive in the SOL region, while it becomes negative in the edge region to balance the radial ion pressure gradient, as imposed by the radial force balance (see \sref{sec:compoedge}).

The estimate given by \eref{eq:ersol} is displayed in \fref{fig:omper} for the SOL. As observed, the radial electric field in the SOL closely follows the radial gradient of $T_{\parallel e}$. The deviations between $E_r$ and $- \partial_r T_{\parallel e} / e$ may stem from the role of the parallel resistivity and finite density gradients, which are neglected in \eref{eq:ersol}. Overall, the transition from \eref{eq:eresimtate} to \eref{eq:ersol} leads to a rapid sign reversal of the electric field across the separatrix, generating strong flow shear and in turns regulate turbulent transport. 

\section{Edge Turbulence Characterization} \label{sec:turbulencecharacteristics}

We now examine the edge turbulence characteristics predicted by the spectral simulations. Our analysis shows that the spatiotemporal Fourier spectra closely match those from GK grid-based simulations \cite{ulbl2023}, with minimal sensitivity to the number of spectral coefficients. A comparison of the temporal Fourier spectrum with linear theory estimates (see \ref{appendix:linearfrequency}) confirms that the turbulence is TEM-dominated.

This section first presents a Fourier decomposition across different flux surfaces (\sref{subsec:fluxsurfacefourier}), followed by a detailed temporal Fourier analysis on a closed flux surface in the edge (\sref{subsec:temporalfourier}).

\subsection{Flux Surface Fourier Analysis of Electrostatic Potential} \label{subsec:fluxsurfacefourier}

We begin by performing a Fourier analysis of the electrostatic potential $\phi_1$ on different flux surfaces, following the same procedure outlined in \cite{ulbl2023}. To this end, we introduce the binormal coordinate $y = r \theta$, where $\theta$ is the geometrical poloidal angle, and $r = L_s / (2\pi)$ is an effective radius associated with the total poloidal arc length $L_s$. The Fourier amplitude of $\phi_1$, denoted by $\hat{\phi}_1(k_y)$ at wavenumber $k_y$, is obtained via the Fourier transform:

\begin{equation} \label{eq:fouriertransform}
    \hat{\phi}_1(k_y) = \int d y \, e^{- i k_y y } \phi_1(y).
\end{equation}

Here, the wavenumber $k_y$ corresponds to a single poloidal mode number $m$, such that $k_y = m / r$. The Fourier transform defined in \eref{eq:fouriertransform} can be applied to any quantity defined on a closed flux surface.

\begin{figure}
    \centering
    \includegraphics[scale=0.5]{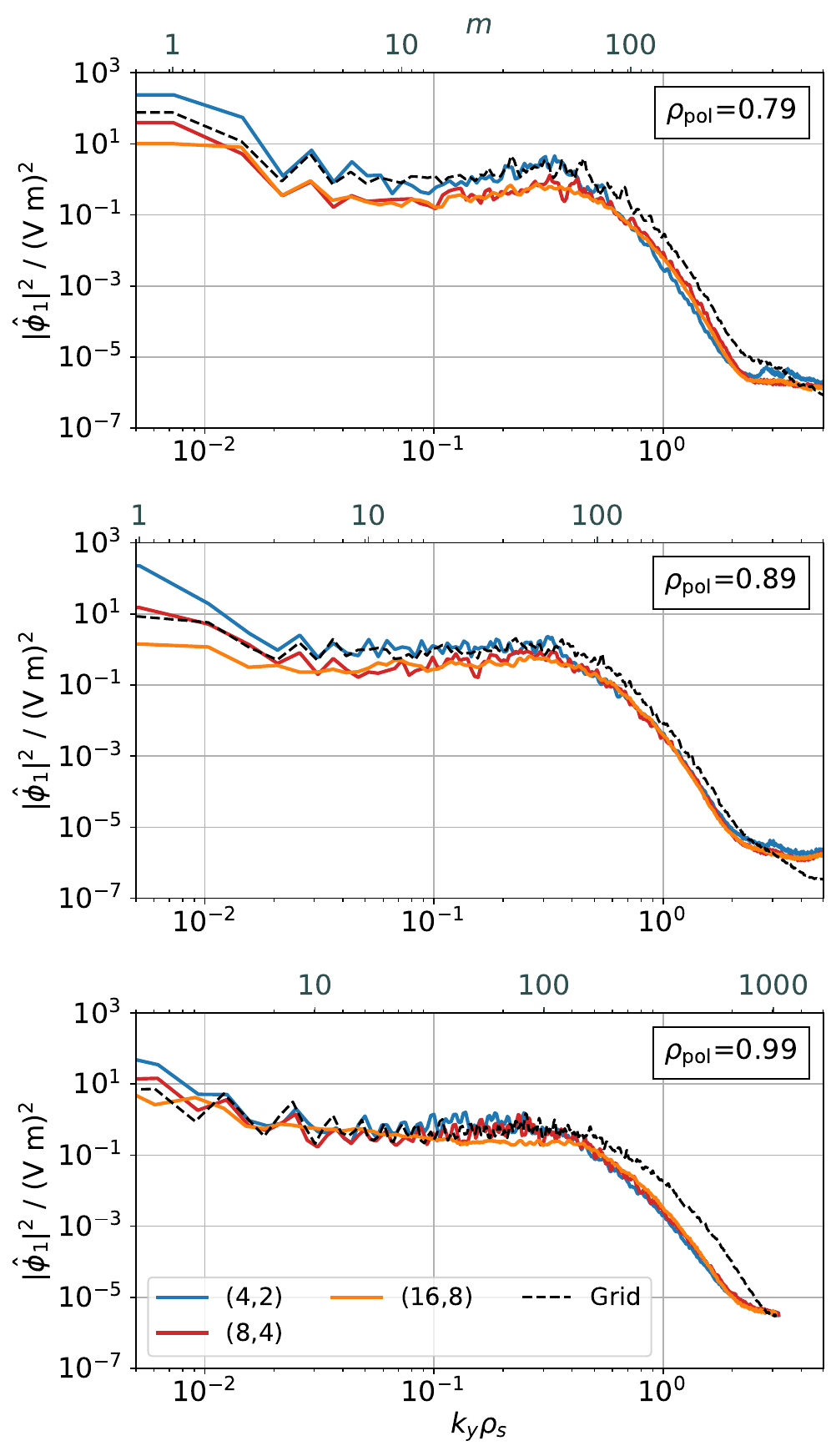}
     \caption{Amplitude of the Fourier component, $\hat \phi_1$, shown as a function of the normalized poloidal wavenumber $k_y \rho_{s}$ (bottom x-axis) and of the poloidal mode number $m$ (top x-axis) on different flux surfaces, $\rho_\mathrm{pol} = 0.79$ (top), $\rho_\mathrm{pol} = 0.89$ (center), and $\rho_\mathrm{pol} = 0.99$ (bottom). The spectra from the spectral and grid-based \cite{ulbl2023} simulations are shown by the solid colored lines and black dashed, respectively and are averaged over the toroidal direction and over $0.1$~ms.}
    \label{fig:espotfourier}
\end{figure}

We compute the spectra of $\hat \phi_1(k_y)$ on the three flux surfaces, with normalized flux surface labels $\rho_{\mathrm{pol}} = 0.79$, $0.89$ and $0.99$, and compare them with the grid-based simulation \cite{ulbl2023}. The results are shown in \fref{fig:espotfourier} where the spectra are plotted as a function of the normalized poloidal wavenumber $k_y \rho_{s}$ (and poloidal mode number $m$ on the top x-axis). Here, the local ion sound Larmor radius, $\rho_s = c \sqrt{ \avgy{T_e} m_i }/ (q_i \avgy{B})$ ($\avgy{\cdot}$ denotes the average over $y$), is computed by averaging the local electron temperature $T_e$ and magnetic field strength $B$ on each flux surface. The spectra are obtained by performing an average over the toroidal direction and over a time window of $0.1$~ms (similarly with \cite{ulbl2023}). The spectra of the electrostatic potential $\phi_1$ are charaterized by a broad turbulent spectrum until $k_y \rho_s \sim 1$ and decay at smaller perpendicular scales. Only minor changes are observed in the different spectra which indicate the similar nature of turbulence across the edge. 

Overall, the Fourier spectra display in \fref{fig:espotfourier} obtained the spectral simulations agree well with the ones of the grid-based simulations and negligible differences are observed between the different spectral resolutions. Specially, a good agreement is observed on length scales $k_y \rho_s \lesssim 0.5$ where the turbulent fluxes typically peak (see below). On the other hand, the small deviations observed at $k_y \rho_s \gtrsim 1$ should not be considered physical as the edge and SOL turbulence GK model (see \sref{sec:edgeandsolturbulentmodel}) is based on the long wavelength approximation and perpendicular hyperdiffusion \cite{frei2024} might play a non-negligible role at these small perpendicular scales. 
Finally, we note that a linearized polarization model is used in \cite{ulbl2023}, whereas the spectral simulations employ a nonlinear polarization in the QN condition (see \eref{eq:poissonvspec}). This discrepancy in the treatment of polarization within the QN condition may contribute to differences observed in the spectra shown in \fref{fig:espotfourier}.

\subsection{Temporal Fourier Analysis} \label{subsec:temporalfourier}

We now conduct a temporal Fourier analysis on a selected flux surface at $\rho_{\mathrm{pol}} = 0.92$ and aim to identify the dominant instability driving the observed turbulence. In order to obtain the temporal Fourier spectra, the Fourier components $\hat \phi_1(k_y)$ are again Fourier transformed but in time. This is done by introducing the temporal Fourier component $\hat \phi_1(k_y, \omega)$ of $\hat \phi_1(k_y)$, which is obtained via 

\begin{equation} \label{eq:temporalfourier}
    \hat \phi_1(k_y, \omega) = \int d t \hat \phi_1(k_y) e^{i \omega t}. 
\end{equation}
Here, the mode frequency $\omega$ can be either negative and positive, depending on the mode propagation direction. Given the direction of the poloidal geometrical angle $\theta$ (oriented counterclockwise in our convention), the mode frequency $\omega$ is defined positive ($\omega > 0$) if the mode propagates in the electron diamagnetic direction such that $\bm k_\perp \cdot \bm V_{e}^* > 0$, where $\bm V_{e}^* = - c \bm B \times \nabla_\perp P_e / (e n_e B^2)$ is the electron diamagnetic velocity. On the other hand, if $\omega < 0$, the mode propagates in the ion diamagnetic direction, i.e. $\bm k_\perp \cdot \bm  V_{e}^* < 0$. To determine the mode propagation direction, it is important to perform the temporal Fourier analysis in the co-moving frame rotating with the plasma due the poloidal velocity, approximated by the $\bm E\times \bm B$ drift. We follow the same procedure as outlined in \cite{ulbl2023} to apply the Doppler shift.

Besides investigating the mode propagation direction to identify the underlying instability, we also compare the mode frequency $\omega$ with predictions derived from linear theory. While accurately identifying these instabilities in global simulations is significantly more challenging than in, e.g., flux-tube analysis \cite{frei2023}, comparison with linear theory still provides valuable physical insights into the turbulent properties. In the present analysis, we focus on linear frequency estimates of electrostatic instabilities due to the low values of $\beta_e$ found in the simulations. \ref{appendix:linearfrequency} reports the analytical expressions of the linear frequency estimates we use below. These are the electron drift wave frequency $\omega_e^*$ in \eref{eq:wdw}, the collisionless and dissipative TEM frequencies, $\omega_{\mathrm{cTEM}}$ in \eref{eq:wctem} and $\omega_{\mathrm{dTEM}}$ in \eref{eq:wdtem}, and, finally, the ITG frequency, $\omega_{i}^*$ in \eref{eq:wsitg}. To compute the linear frequencies, local parameters (e.g., local temperature and equilibrium density and temperature gradients) are calculated by an average over the flux surface in the poloidal plane and over $0.1$~ms.

\begin{figure}
    \centering
    \includegraphics[scale=0.45]{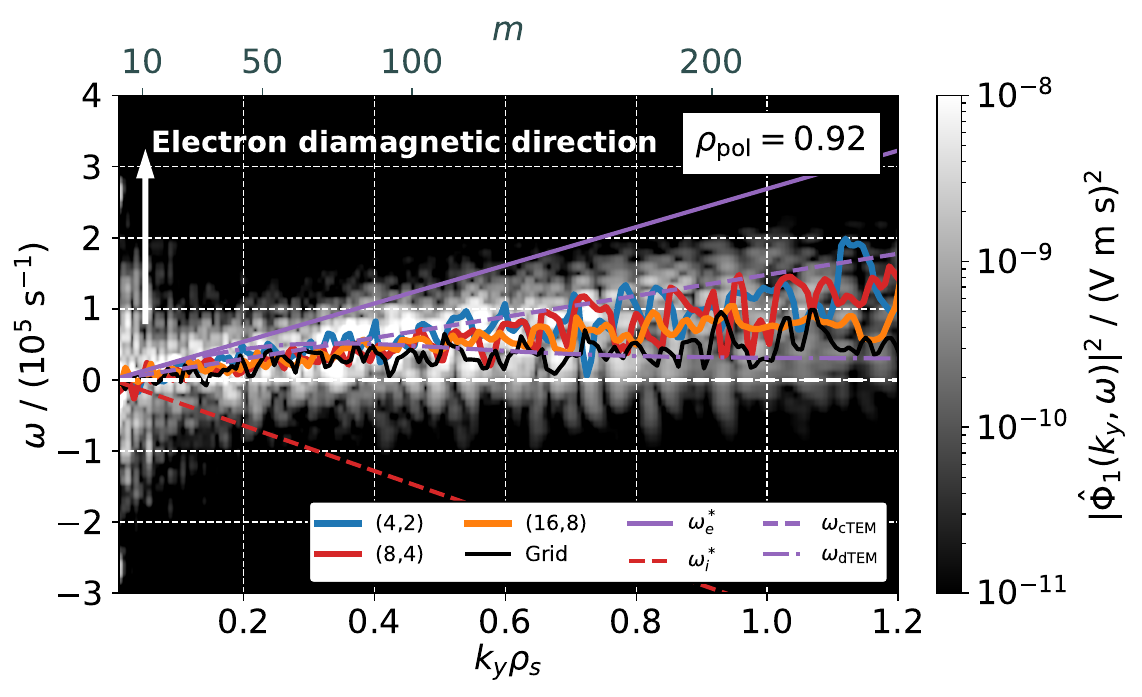}
    \caption{Temporal Fourier spectra of $| \hat \phi_1(k_y, \omega)|^2$ computed on the flux surface $\rho_\mathrm{pol} = 0.92$ obtained from the $(16,8)$ spectral simulation. The spectrum maxima (calculated using a rolling average) obtained in the $(4,2)$ and $(8,4)$ spectral \cite{frei2024} and grid-based \cite{ulbl2023} simulations are shown by the solid colored and black lines, respectively. The linear frequency estimates (defined in \sref{appendix:linearfrequency}) are also shown for comparison. The analysis is performed on a single poloidal plane and the spectra are averaged over $0.1$~ms at steady-state.}
    \label{fig:fouriertemporal}
\end{figure}

\Fref{fig:fouriertemporal} displays the temporal Fourier spectrum (averaged over the toroidal direction and over $0.1$~ms) from the $(16,8)$ spectral simulation on the $\rho_\mathrm{pol} = 0.92$ flux surface. For comparison, the positions of the spectral maxima obtained from the $(4,2)$ and $(8,4)$ spectral simulations, as well as grid-based simulations, are indicated by solid colored lines. The spectrum exhibits a broadband nature, with most turbulent structures propagating in the electron diamagnetic direction ($\omega > 0$). This corresponds to the counterclockwise direction in the poloidal plane (with the magnetic field pointing in the perpendicular plane). The Fourier components are the largest near $k_y \rho_s \sim 0.4$ with a mode frequency in the range of $0.5$~MHz, consistent with the findings of \cite{ulbl2023}. The position of the maxima demonstrate excellent agreement over the range of spatial scales driving transport (see next section), extending up to $k_y \rho_s \sim 0.8$. At larger $k_y \rho_s$, deviations become noticeable most likely due to hyperdiffusion effects and statistical noise. Overall, the amplitudes of the Fourier components are very similar to those displayed in Fig. 13 of  \cite{ulbl2023}, confirming that the spectral approach captures the same temporal and spatial turbulent characteristics to those in grid-based simulations. 

We remark that the spectra from the $(4,2)$ and $(8,4)$ simulations are very similar to the spectral of the $(16,8)$ simulation shown in \fref{fig:fouriertemporal}. This suggests that the typical frequencies of the turbulent structures are weakly affected by the number of spectral coefficients retained in the simulations. This observation aligns with findings from local flux-tube simulations based on a similar approach \cite{frei2023}, where the real frequencies of microinstabilities (e.g., TEM) have been shown to be weakly affected by the spectral resolution in contrast to their linear growth rates. The sensitivity to spectral resolution was also found to increase with the binormal wavenumber $k_y$, which is also consistent with the trend shown in  \fref{fig:fouriertemporal}.

Evidence of TEM-driven turbulence can be inferred from \fref{fig:fouriertemporal} by examining the direction and the amplitude of the linear frequency estimates (also shown in \fref{fig:fouriertemporal}) relative to the spectral peak. The propagation direction can be inferred from the ITG linear frequency $\omega_i^*$ (defined negative the ion diamagnetic direction), which reveals that the turbulence propagates in the electron diamagnetic direction. The electron drift wave frequency, $\omega_e^*$, which scales with the density gradient, yields frequencies that are too large (of the order of $0.1$~MHz). However, a match better agreement is found with the cTEM and dTEM frequencies, given by \eref{eq:wctem} and \eref{eq:wdtem}. Notably, for $k_y \rho_s \lesssim 0.3$, both spectral and grid-based simulations closely follow $\omega_{\mathrm{cTEM}}$. As also noted in \cite{ulbl2023}, deviations from the linear $k_y \rho_s$ scaling of $\omega_{\mathrm{cTEM}}$ (see \eref{eq:wctem}) appear for $k_y \rho_s \gtrsim 0.6$. Interestingly, at shorter wavelengths, the maxima align well with the dTEM frequency, $\omega_{\mathrm{dTEM}}$, which follows a $\sim (k_y \rho_s)^{-2}$ 
scaling at large $k_y$ \cite{connor2006}. Therefore, dTEMs might be present in our simulations and could explain the departure from linear scaling observed in the temporal Fourier spectra. We remark that this interpretation is consistent with the fact that dTEMs can be unstable at shorter wavelengths as they are strongly stabilised at longer wavelengths when collisionality and density gradients fall below a critical threshold. We also note that similar observations are made on different flux surfaces, indicating that TEM-driven turbulence dominates across the edge. 

Overall, the present analysis demonstrates that the spatial and temporal structures of edge fluctuations in the spectral simulations closely resemble those observed in grid-based simulations, with only a weak dependence on spectral resolution.

\section{Edge Transport Analysis} \label{sec:turbulenttranapsortanalysis}

Having studied the edge turbulence properties, we now assess its consistency in predicting particle and heat transport, which are important for predictive modelling. We begin with a Fourier analysis of the turbulent particle and heat fluxes in \sref{subsec:fourierspectrafluxes}, comparing their spectra and phase shifts with those from grid-based simulations \cite{ulbl2023}. Our results show strong agreement across all transport channels. In \sref{subsec:powerbalance}, we examine the total power crossing the separatrix. Then, in \sref{subsec:divertor}, we analyze the heat flux at the divertor and measure the associated falloff length. Finally, in \sref{sec:diamagnetictransport}, we evaluate the contributions of diamagnetic (neoclassical) particle and heat fluxes and show that they can be neglected in the present case. 

\subsection{Analysis of Particle and Heat Fluxes} \label{subsec:fourierspectrafluxes}

Building on the turbulence characterization from the previous section, we perform a Fourier analysis of the radial particle and heat turbulent fluxes in the edge. We focus on the fluxes driven by turbulent electrostatic $\bm{E} \times \bm{B}$ drifts, which is a reliable proxy for the total particle and heat transport. Indeed, the contributions from the diamagnetic fluxes, associated with magnetic drifts (\sref{sec:diamagnetictransport}), as well as flutter-induced transport, can be neglected, as they remain small and negligible in the present low-$\beta$ L-mode discharge.

Therefore, the total radial particle flux, denoted by $\Gamma_\alpha$, can be written as 

\begin{eqnarray} \label{eq:gammaalpha}
    \Gamma_{\alpha}  =  \int d W f_\alpha \dot \bm R \cdot \nabla r  \simeq    n_\alpha  \frac{c\bm B \times \nabla \phi_1}{B^2} \cdot \nabla r.
\end{eqnarray}
Here, the radial projection of the total drift, $\dot \bm R \cdot \nabla r$, is approximated by the radial projection of the electrostatic $\bm{E} \times \bm{B}$ velocity, i.e. $ c\bm B \times \nabla \phi_1 / B^2 \cdot \nabla r$. To analyze the particle flux $\Gamma_{\alpha}$ \eref{eq:gammaalpha}, we apply the Fourier transform defined in \eref{eq:fouriertransform}. By introducing the Fourier components $\hat n_\alpha(k_y)$ of the plasma density and $\hat \phi_1(k_y)$ of the electrostatic potential, the Fourier component $\hat \Gamma_\alpha(k_y)$ of the particle flux can be expressed as \cite{ulbl2023}

\begin{eqnarray} \label{eq:gammaesfourier}
    \hat \Gamma_\alpha(k_y) & = \frac{1}{\avgy{B}} \mathrm{Re}\left( i k_y \hat n_\alpha(k_y) \hat \phi_1^*(k_y) \right) \nonumber \\
    & =  k_y \left|\hat n_\alpha(k_y) \right| \left|\hat \phi_1(k_y) \right| \nonumber \\
     & \times \sin \alpha\left(\hat n_\alpha(k_y) ,  \hat \phi_1(k_y)\right) \avgy{B}^{-1},
\end{eqnarray}
where $\alpha( \hat n_\alpha(k_y), \hat \phi_1(k_y) )$ is the phase shift between $\hat n_\alpha(k_y)$ and $\hat \phi_1 (k_y)$ and is defined by $ \alpha\left
( \hat n_\alpha(k_y), \hat \phi_1(k_y) \right)
= \mathrm{Im} \left( \log\left( \hat n_\alpha^*(k_y)\hat \phi_1(k_y) \right) \right)$. The phase shift contains important information about the direction and amplitude of the particle fluxes, as discussed below.

Similarly to $\Gamma_\alpha$, the total radial heat flux, $Q_\alpha$, is expressed as

\begin{eqnarray} \label{eq:qalpha}
    Q_\alpha  & =   \frac{ m_\alpha }{2}  \int d W   v^2 f_\alpha  \dot \bm R \cdot \nabla r \nonumber \\
    & \simeq \frac{ m_\alpha }{2}  \int d W   v^2 f_\alpha  \frac{c\bm B \times \nabla \phi_1}{B^2} \cdot \nabla r,
\end{eqnarray}
where the magnetic drifts and flutter in $\dot \bm R \cdot \nabla r$ are also neglected as in \eref{eq:gammaalpha}. Similarly to \eref{eq:gammaesfourier}, we Fourier analyze \eref{eq:qalpha} by introducing the Fourier component, $\hat Q_\alpha(k_y)$, given by

\begin{eqnarray} \label{eq:qesfourier}
    \hat Q_\alpha(k_y)  & = \frac{3}{2} \frac{k_y}{\avgy{B}} \mathrm{Re} \left( i \avgy{T_\alpha} \hat n_\alpha(k_y) \hat \phi_1^*(k_y)  \right. \nonumber 
    \\ & \left.+ i \avgy{n_\alpha} \hat T_\alpha(k_y) \hat \phi_1^*(k_y) \right) \nonumber \\
    & = \hat Q_\alpha^{\mathrm{conv}}(k_y)  + \hat Q_\alpha^{\mathrm{cond}}(k_y).
\end{eqnarray}
In \eref{eq:qesfourier}, $\hat Q_\alpha$ is splitted into a convective and conductive parts, $\hat Q_\alpha^{\mathrm{conv}}$ and $\hat Q_\alpha^{\mathrm{cond}}$ respectively. These are defined by

\begin{eqnarray} \label{eq:qconv}
\hat Q_\alpha^{\mathrm{conv}}(k_y) & = \frac{3}{2} \frac{\avgy{T_\alpha} k_y}{\avgy{B}} \left|\hat n_\alpha(k_y) \right| \left|\hat \phi_1(k_y) \right|  \nonumber \\
& \times \sin \alpha\left(\hat n_\alpha(k_y) ,  \hat \phi_1(k_y)\right),   \\ 
\hat Q_\alpha^{\mathrm{cond}}(k_y)  & = \frac{3}{2}   \frac{\avgy{n_\alpha} k_y}{\avgy{B}} \left|\hat T_{ \alpha}(k_y) \right| \left|\hat \phi_1(k_y) \right|  \nonumber \\
& \times \sin \alpha\left(\hat T_{ \alpha}(k_y) ,  \hat \phi_1(k_y)\right), \label{eq:qcond}
\end{eqnarray}
respectively. Here, $\alpha\left(\hat T_{ \alpha}(k_y) ,  \hat \phi_1(k_y)\right)$ is the phase shift between the total temperature $T_\alpha = (T_{\parallel \alpha} + 2 T_\perp) /3$ and the electrostatic potential $\phi_1$.

From \eref{eq:qconv}, it can be observed that the convective part of the total heat flux $\hat{Q}_\alpha$ is proportional to the particle flux such that $\hat Q_\alpha^{\mathrm{conv}} = 3\hat \Gamma_\alpha \avgy{T_\alpha} / 2$ and thus depends on the density fluctuations, while the conductive part, given in \eref{eq:qcond}, depends on the temperature fluctuations \cite{zholobenko2023}. In addition, the Fourier decompositions given in \eref{eq:gammaesfourier} and \eref{eq:qesfourier} show that the amplitudes and directions (inwards or outwards) of the radial fluxes are also related to the phase shifts between $\hat  \phi_1$ and the density and temperature fluctuations. More precisely, the fluxes are maximal if $\alpha = \pm \pi / 2$ (for, e.g., interchange-like mode) and while they vanish if $\alpha = n \pi$, with $n$ an integer (e.g., drift-wave modes). Positive (negative) phase shifts indicate that the transport is outwards (inwards). Therefore, important insights on the particle and heat turbulent transport can be inferred from the phase shift $\alpha(\cdot, \hat \phi_1)$, which we present below.

\begin{figure}
    \centering
    \includegraphics[scale=0.23]{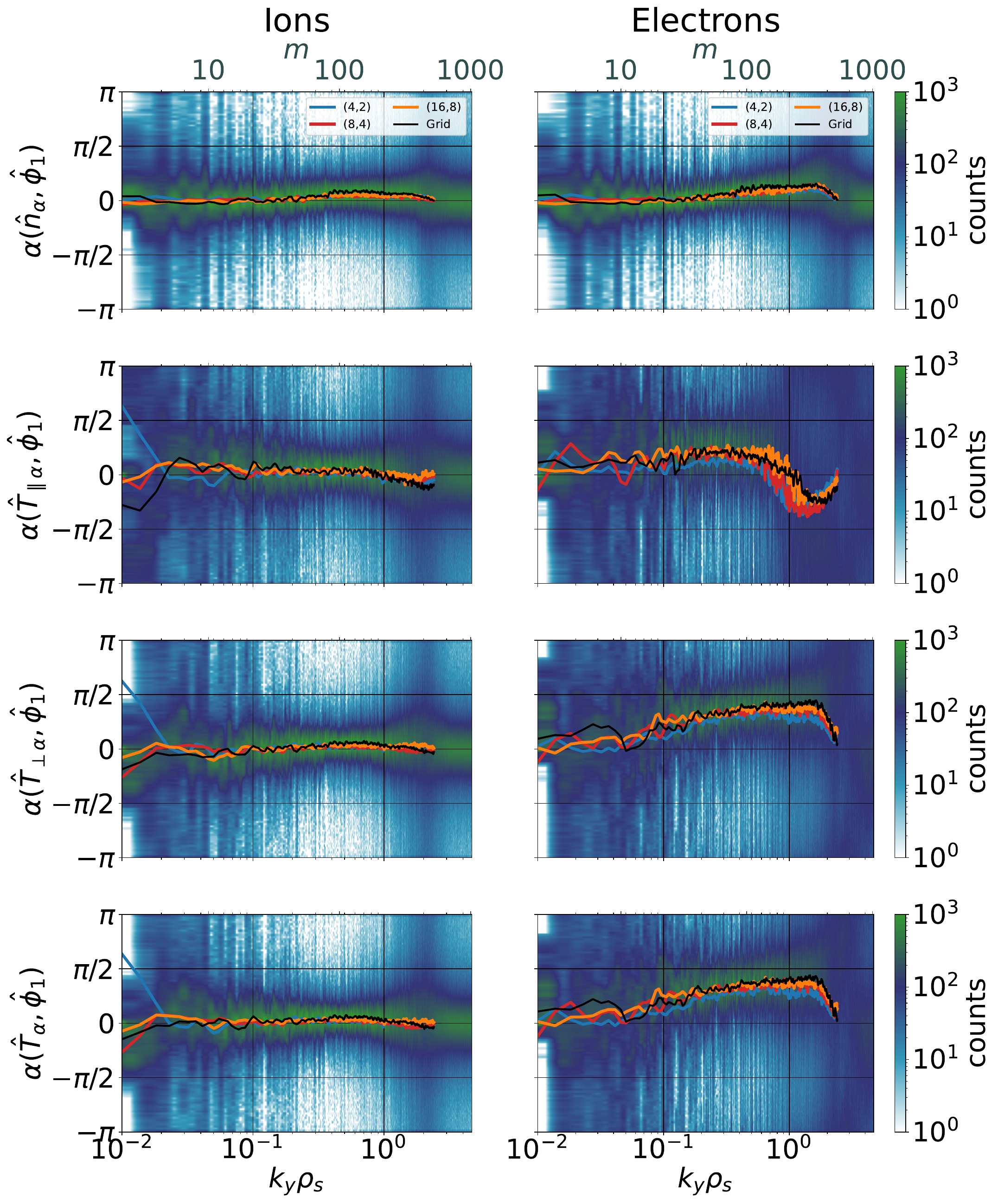}
    \caption{Histograms of the phase shifts, $\alpha(\cdot, \hat{\phi}_1)$, of (from top to bottom) the density $n_\alpha$, parallel temperature $T_{\parallel \alpha}$, perpendicular temperature $T_{\perp \alpha}$, and total temperature $T_\alpha$, with the electrostatic potential $\phi_1$ as a function of the normalized poloidal wavenumber $k_y \rho_s$ (bottom axis) and poloidal mode number $m$ (top axis). The phase shifts are computed from the $(16,8)$ spectral simulations for ions (left column) and electrons (right column). The peak positions of $\alpha(\cdot, \hat{\phi}_1)$ are shown for both spectral (colored solid lines) and grid-based (black solid lines) simulations for comparison. The counts are computed over the toroidal direction and a time window of $0.2$~ms to ensure good statistics.}
    \label{fig:phaseshift}
\end{figure}

\Fref{fig:phaseshift} presents the histograms of the phase shifts on the $\rho_{\mathrm{pol}} = 0.92$ flux surface. The counts are computed over the toroidal direction and a time interval of $0.2$~ms for good statistics. While the figure explicitly displays results from the spectral $(16,8)$ simulation, the histogram peaks from the other spectral and grid-based simulations are also included for comparison. In addition, we remark that the histograms from the other spectral simulations are similar, qualitatively and quantitatively, to \fref{fig:phaseshift}, and thus are not shown explicitly. While all ion phase shifts remain close to zero, the electron temperature phase shift increases almost linearly with $k_y$, reaching a peak between $\pi/4$ and $\pi/2$ near $k_y \rho_s \sim 0.5$. In particular, the phase shift between the electron perpendicular temperature and the electrostatic potential, $\alpha(\hat{T}_{\perp e}, \hat{\phi}_1)$, exhibits a clear increase with $k_y$, peaking near $\pi/2$ for $k_y \rho_s \sim 0.5-0.9$, which corresponds to the range of $k_y$ values where the temporal Fourier spectrum peaks (see \fref{fig:fouriertemporal}). In contrast, $\alpha(\hat{T}_{\parallel e}, \hat{\phi}_1)$ remains below $\pi/4$. This highlights the dominant role of perpendicular temperature fluctuations in the conductive heat flux $Q_e^{\mathrm{cond}}$, an observation consistent with TEM-dominated turbulence \cite{dannert2005}. We also observe that all the phase shifts are positive, indicating outward transport of particles and heat.

 Overall, the phase shift analysis demonstrates strong agreement across all transport channels between the spectral and grid-based simulations. Furthermore, this agreement improves with increased spectral resolution, especially at high $k_y \rho_s$ in the electron channels. A direct comparison of \fref{fig:phaseshift} with Fig. 15 of \cite{ulbl2023} can also confirm the findings of this section.
 
\begin{figure*}
    \centering
    \includegraphics[scale=0.45]{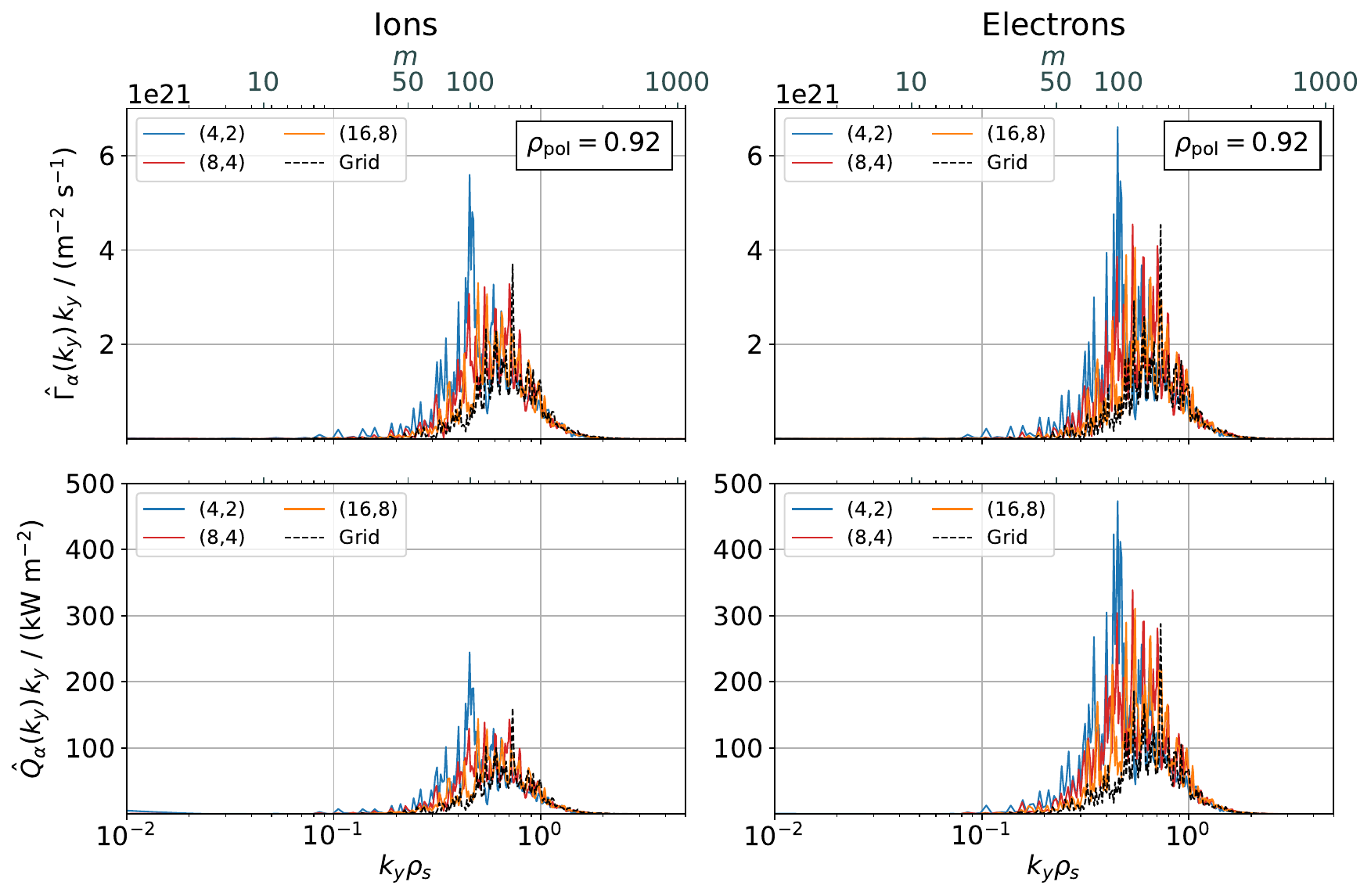}
     \caption{Fourier spectra of the turbulent radial particle flux $\Gamma_\alpha$ (top) and heat flux $Q_\alpha$ (bottom) calculated on the $\rho_\mathrm{pol} = 0.92$ flux surface for ions (left column) and electrons (right column). The spectra are plotted as functions of the normalized poloidal wavenumber $k_y \rho_s$ (bottom axis) and the poloidal mode number $m$ (top axis). The results are shown for both spectral (coloured solid lines) and grid-based (dashed black lines) simulations. The spectra are averaged in the toroidal direction and over $0.2$~ms.}
    \label{fig:fluxfourier}
\end{figure*}

From the phase shifts, the Fourier spectra of the radial particle and heat fluxes, $\hat{\Gamma}_\alpha$ and $\hat{Q}_\alpha$ in \eref{eq:gammaesfourier} and \eref{eq:qesfourier} respectively, are shown in \fref{fig:fluxfourier}. The good agreement between the spectral and grid-based simulations is further confirmed. More precisely, this agreement improves with increasing spectral resolution, particularly at low $k_y \rho_s$. The residual discrepancies observed at short wavelengths ($k_y \rho_s \gtrsim 0.8$) are attributed to numerical errors in the interpolation process and statistical average. From \fref{fig:fluxfourier}, the spectral approach successfully reproduces both the position and shape of the turbulent spectrum observed in the grid-based simulations.  

From the Fourier spectra depicted in \fref{fig:fluxfourier}, we observe that the $k_y \rho_s \sim 0.5$ components primarily drive the turbulent transport, consistent with the phase shifts analysis, i.e. $\alpha(\cdot, \hat \phi_1) \sim \pi /2$ at this location and presented in \fref{fig:phaseshift}. While the ion and electron turbulent particle fluxes exhibit similar spectra (influenced by the particle balance in steady-state), the Fourier components of the turbulent ion heat flux, $\hat Q_i$, exhibit smaller amplitudes compared to $\hat Q_e$. This observation supports the predominance of electrons in carrying the heat flux, in accordance with the TEM-dominated turbulence hypothesis.

\begin{figure*}
    \centering
    \includegraphics[scale=0.45]{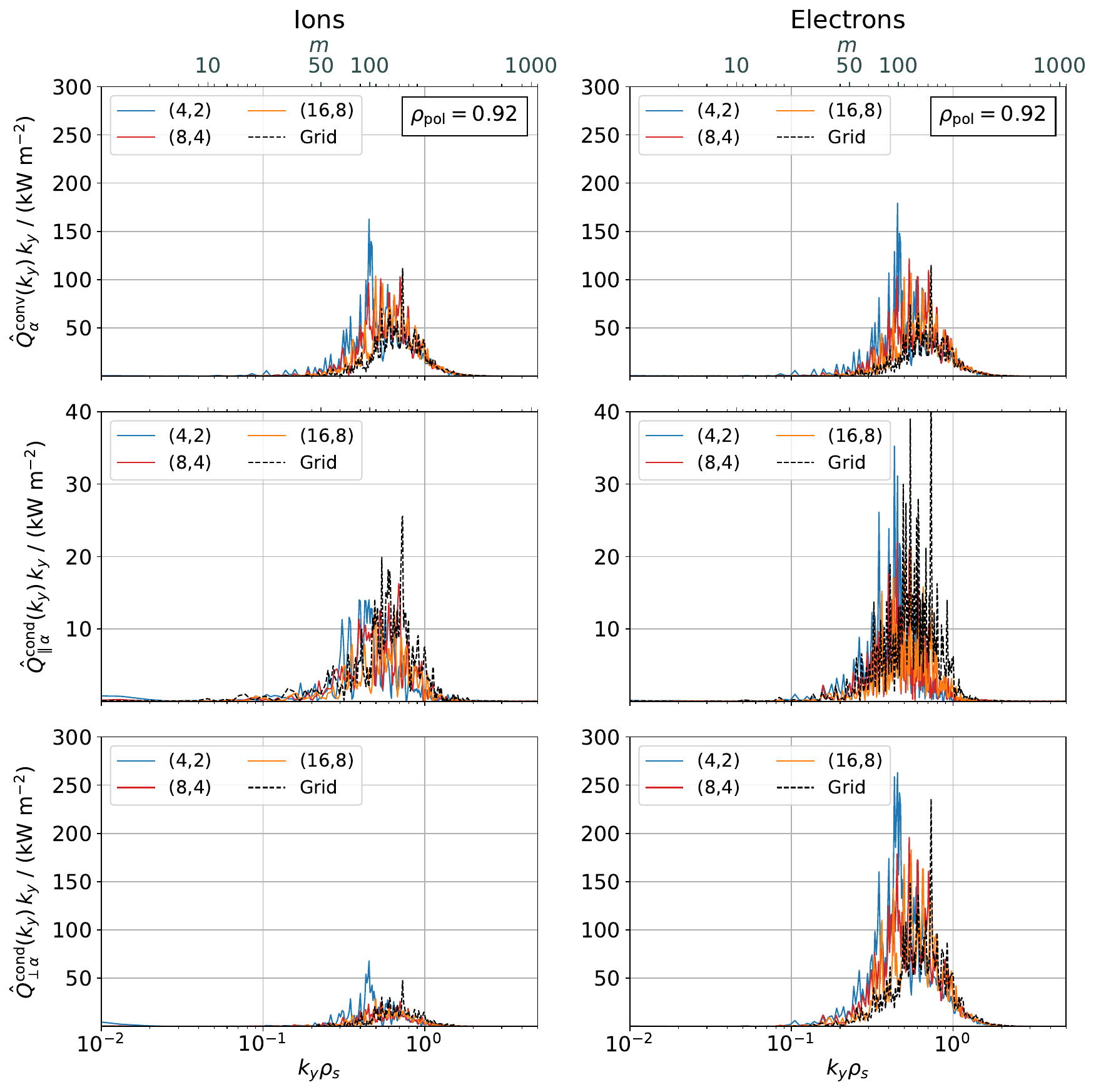}
     \caption{Fourier spectra calculated on the $\rho_\mathrm{pol} = 0.92$ flux-surface of the convective part $Q_\alpha^{\mathrm{conv}}$ (top row), the parallel $Q_{ \parallel \alpha}^{\mathrm{cond}}$ (middle row) and perpendicular $Q_{ \perp \alpha}^{\mathrm{cond}}$ (bottom row) components of the conductive part $Q_\alpha^{\mathrm{cond}}$ for the ions (left column) and electrons (right column). Here, the spectral and grid-based simulations are shown by the colored solid and dashed black lines. }
\label{fig:fluxfourierqconvcond}
\end{figure*}

The composition of the radial heat flux can be further analyzed by examining the contributions of the convective and conductive components. In particular, the conductive heat flux, defined in \eref{eq:qcond}, can be decomposed into parallel and perpendicular parts based on the temperature relation $\hat{T}_\alpha = (\hat{T}_{\parallel \alpha} + 2 \hat{T}_{\perp \alpha})/3$. Thus, uing \eref{eq:qcond}, the conductive heat flux $\hat{Q}_\alpha^{\mathrm{cond}}$ can be written as 

\begin{equation} \label{eq:qcondparperp}
    \hat{Q}_\alpha^{\mathrm{cond}}(k_y)  =  \hat{Q}_{\parallel \alpha}^{\mathrm{cond}}(k_y) +  \hat{Q}_{\perp \alpha}^{\mathrm{cond}}(k_y).
\end{equation}  
\Fref{fig:fluxfourierqconvcond} shows the Fourier spectra of the convective, parallel, and perpendicular conductive heat fluxes for both ions and electrons. As observed, the spectral simulations reproduce the shape and the peak position of the spectra for both the convective and conductive heat fluxes. Furthermore, while the convective heat flux is the dominant component for ions, the perpendicular conductive heat flux is the largest contributor to the electron heat flux, highlighting the role of TEM-driven perpendicular temperature fluctuations.

\subsection{Power Across the Separatrix} \label{subsec:powerbalance}

The total power crossing the separatrix and driven by turbulent transport, denoted by $ P_{\mathrm{sep}} $, is a crucial physical quantity in transport predictions. Therefore, to calculate $ P_{\mathrm{sep}} $, we sum the surface-integrated radial turbulent heat flux $Q_\alpha$ over all species, such that

\begin{equation} \label{eq:psep}
P_{\mathrm{sep}}= \sum_\alpha P_\alpha,
\end{equation}  
where  

\begin{equation} \label{eq:poweralpha}
 P_\alpha = \oint dS \avgt{Q_\alpha} \simeq \oint_0^{2\pi} d\phi \oint_0^{L_s} d\ell R(\ell) \avgt{Q_\alpha}.
\end{equation}  
Here, $ \ell $ represents the poloidal arc length of the flux surface near the separatrix and $ \avgt{Q_\alpha} $ is the time-averaged radial heat flux associated with turbulent $\bm{E} \times \bm{B}$ transport. Again here, we neglect the contributions from magnetic drifts and flutter and show that they are indeed small in the next section. We compute $ P_\alpha $ using \eref{eq:poweralpha} on a flux surface in the edge and in the vicinity of the separatrix at $\rho_{\mathrm{pol}} = 0.99$ for both electrons and ions, averaging over a $0.1$~ms time window in a quasi-steady state \cite{ulbl2023}. Note that the total power remains nearly constant (in the statistical sense) across the edge under quasi-steady state conditions because of the absence of localized heat sources.

\Tref{tab:psep} compares the electron ($P_e$), ion ($P_i$), and total ($P_{\mathrm{sep}}$) power measured in the spectral and grid-based simulations. As observed, the spectral simulations exhibit good quantitative agreement with the grid-based predictions. While all spectral predictions remain within $15\%$ of the grid-based results, the best agreement is achieved at a $(16,8)$ velocity-space resolution. Overall, a slight systematic overestimation of $P_e$ is observed, which is consistent with the trends present in the Fourier spectra of the turbulent fluxes as discussed in \sref{subsec:fourierspectrafluxes} and shown in \fref{fig:fluxfourier}. In contrast, the ion power $P_i$ (smaller than $P_e$ due to TEM-dominated turbulence) shows much better agreement with the grid-based predictions, with deviations within $1\%$. This suggests that the ion transport channel is less sensitive to spectral resolution compared to the electron channel.  

Both the spectral and grid-based simulations successfully reproduce the experimental TCV measurement of $ P_{\mathrm{sep}} = 120 $~kW \cite{oliveira2022} within $10\%$. This highlights the reliability of the spectral approach in predicting turbulence and associated transport in the edge with the accuracy of high-fidelity GK simulations. 

\begin{table}
\caption{\label{tab:psep}Power crossing the separatrix ($\rho_{\mathrm{pol}} = 0.99$) for electrons ($P_e$), ions ($P_i$), and total power ($P_{\mathrm{sep}}$) obtained from the spectral \cite{frei2024} and grid-based \cite{ulbl2023} simulations. The powers are averaged over $0.1$~ms in a quasi-steady state. The experimentally measured power is $P_{\mathrm{sep}} \simeq 120$~kW in from TCV \cite{oliveira2022}.}
\begin{indented}
\item[]\begin{tabular}{@{}lllll}
\br
$(N_{v_\parallel}, N_{\mu})$ & Type & $P_e$ / kW  & $P_i$ / kW  & $P_\mathrm{sep}$  / kW  \\
\mr
(4,2) & Spectral & $95.8$ & $46.6$ & $142.4$ \\
(8,4) & Spectral & $98.8$ & $46.1$ & $144.9$ \\
(16,8) & Spectral & $87.3$ & $44.4$ & $131.7$ \\
(80,24) &  Grid & $79.0$ & $46.2$ & $125.2$ \\
\br
\end{tabular}
\end{indented}
\end{table}

\subsection{Divertor Heat Flux and SOL Falloff Length}
\label{subsec:divertor}

Complementary to the previous section, we now analyze the heat flux at the divertor plates and associated SOL falloff length and compare them with \cite{ulbl2023}. We remark that these predictions should not be considered fully reliable due to the currently unrealistic boundary conditions near the divertor plate. Additionally, the herein comparison is affected by different numerical sensitivities to the boundary conditions (despite their physical equivalence), restricting therefore the present comparison to a qualitative level.

Our analysis focuses on the right divertor (located on the low field side) where the largest fraction of heat is deposited \cite{ulbl2023}. The parallel heat flux, projected onto the divertor plate, is expressed by

\begin{equation} \label{eq:qdiv}
    q_{\parallel \alpha}^{\mathrm{div}} = \oint d \phi \avgt{q_{\parallel \alpha}(l)} \sin \alpha_I(l),
\end{equation}
where the parallel heat flux is \cite{ulbl2023,frei2024} given by 

\begin{eqnarray}
q_{\parallel \alpha} &= \int d W v_\parallel f_\alpha \left( \frac{m_\alpha v_\parallel^2}{2} + \mu B \right) \nonumber \\
& = \sqrt{\frac{\tau_\alpha^3}{m_\alpha}} \left(  \sqrt{\frac{3}{2}} \mathcal{N}_\alpha^{30}  + \frac{5}{2} \mathcal{N}_\alpha^{10}   - \mathcal{N}_\alpha^{11} \right).
\end{eqnarray}
In \eref{eq:qdiv}, $l$ defines the poloidal arc-length parametrizing the divertor plates and $\sin \alpha_I(l)$ represents the correction from the fact that the magnetic field lines intercept the divertor plates with a shallow incidence angle (of around $2.5°$). To minimize the effect of the boundary conditions, \eref{eq:qdiv} is calculated on a plane slightly away from the divertor plates which is constructed by tracing back the points located on the divertor by a few toroidal planes. Similarly to \sref{subsec:powerbalance}, the divertor heat flux is averaged over a time period of $0.1$~ms in quasi-steady state.

\begin{figure}
    \centering
    \includegraphics[scale=0.4]{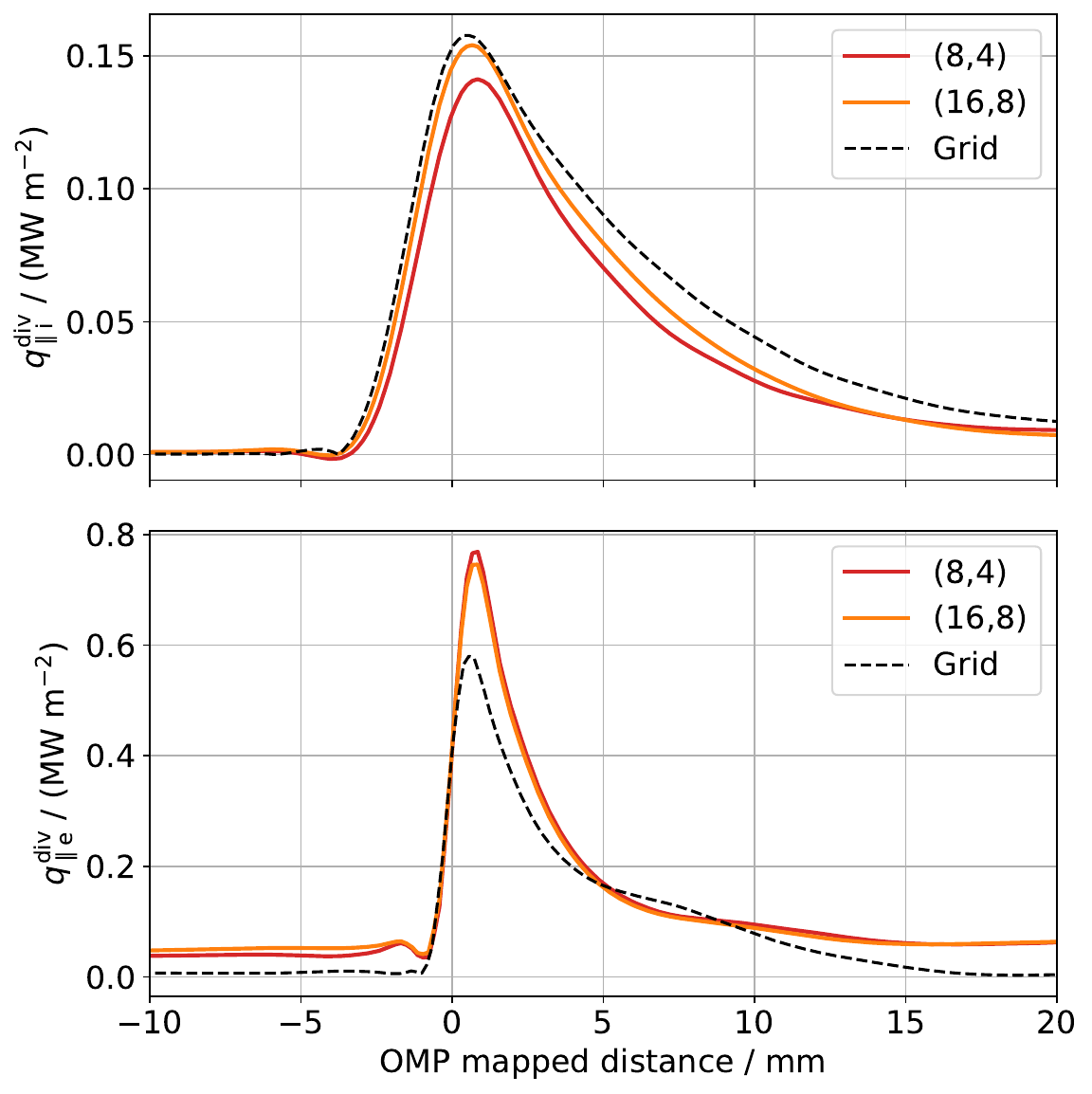}
    \caption{Right divertor heat flux $q_{\parallel \alpha}^{\mathrm{div}} $ (defined in \eref{eq:qdiv}) for ions (top) and electrons (bottom) as a function of the OMP-mapped distance from the separatrix obtained from the spectral (solid colored lines) and grid-based (dashed back lines) simulations. }
    \label{fig:divq}
\end{figure}

\Fref{fig:divq} shows ion and electron heat flux profiles (measured the right divertor) plotted as a function of the OMP-mapped distance from the separatrix \cite{ulbl2023}. The ion divertor heat flux from the spectral approach agrees well with grid-based results, particularly in the $(16,8)$ case, though it is slightly narrower. For electron heat flux, deviations from grid-based predictions are larger. The peak value is overestimated by about $15\%$, and the profile is narrower. Despite these minor deviations, the divertor heat flux is in reasonable agreement with \cite{ulbl2023}. The $(4,2)$ case is excluded due to spurious oscillations in $q_{\parallel \alpha}$, which affect the heat flux profile, likely due to the sensitivity of the flux to the spectral coefficient $\mathcal{N}_\alpha^{30}$ at $(4,2)$.

Another physically relevant quantity is the SOL falloff length, $\lambda_q$ \cite{eich2013,chang2017}. This can be measured by fitting the total divertor heat flux, $q_\parallel^{\mathrm{div}} = \sum_\alpha q_{ \alpha \parallel}^{\mathrm{div}}$, with the empirical Eich fit function \cite{eich2013}. We follow the same fitting procedure as described in \cite{ulbl2023} and apply the Eich fit to $q_\parallel^{\mathrm{div}}$ obtained from the spectral simulations. As a result, the SOL falloff length is found to be $\lambda_q \simeq (2.9 \pm 0.06)$~mm for the $(16,8)$ case. A similar result is obtained for the $(8,4)$ case. In comparison to \cite{ulbl2023}, which reports $\lambda_q \simeq 3.7$~mm, the SOL falloff length predicted by the spectral simulations are approximately $20\%$ narrower. This deviation can be attributed to the narrower electron divertor heat flux depicted in \fref{fig:divq}. Despite discrepancies in the SOL falloff length, enhancing the accuracy of the SOL description (neutrals and sheath boundary conditions) is expected to significantly improve the current comparison and lead to better accuracy of the results.

\subsection{Analysis of Diamagnetic Fluxes
} \label{sec:diamagnetictransport}

While the total particle and heat transports are shown to be entirely dominated by the turbulent $\bm{E} \times \bm{B}$ velocity, it is worth examining the contributions of the diamagnetic (neoclassical) fluxes to $\Gamma_\alpha$ and $Q_\alpha$ as they may play a significant role in high-performance scenarios \cite{zholobenko2024}. The present analysis is focused only on the $(8,4)$ spectral simulation since the diamagnetic fluxes were only diagnosed for this case. The fluid limit of the diamagnetic fluxes (associated with a Maxwellian distribution function) is also investigated to illustrate the importance of kinetic modifications.

The diamagnetic particle and heat fluxes, $\Gamma_{D \alpha}$ and $Q_{D \alpha}$ respectively, are the fluxes driven by the radial projection of the velocity-space dependent magnetic drift, denoted by $\bm V_{D\alpha}$, being the sum of the $\nabla B$ and curvature drifts. From the equation of motion given in \eref{eq:dotR}, $\bm V_{D\alpha}$ can be written as

\begin{equation} \label{eq:vdia}
    \bm V_{D\alpha} = \frac{c v_\parallel^2 m_\alpha}{q_\alpha B}\nabla  \times \bm b + \frac{c \mu}{q_\alpha B}\bm b \times \nabla B,
\end{equation}
with $B_\parallel^* \simeq B$ in the denominator. 

Using \eref{eq:vdia} and noticing the velocity-space dependence, the radial diamagnetic particle flux, $\Gamma_{D \alpha}$, is expressed as

\begin{eqnarray} 
 \Gamma_{D \alpha} &= \int d W f_\alpha \bm V_{D \alpha}\cdot \nabla r \nonumber\\
 & =  \frac{c   P_{\parallel \alpha}}{q_\alpha B} \nabla \times \bm b \cdot \nabla r  + \frac{c  P_{\perp \alpha}}{q_\alpha B^2}  \bm b \times \nabla B \cdot \nabla r. \label{eq:gammaD}
\end{eqnarray}

Similarly, the radial diamagnetic heat flux, $Q_{D \alpha}$, is 

\begin{eqnarray} 
 Q_{ D\alpha} &= \int d W \frac{1}{2} m_\alpha v^2 f_\alpha \bm V_{D \alpha}\cdot \nabla r \nonumber \\
 & =  \frac{c K_{\parallel \alpha}}{q_\alpha B } \nabla \times \bm b \cdot \nabla r  \nonumber \\
 & +  \frac{c K_{\perp \alpha}}{q_\alpha B^2 }  \bm b \times \nabla B \cdot \nabla r   , \label{eq:qD}
\end{eqnarray}
where we introduce the quantities $K_{\parallel \alpha} $ and $ K_{\perp \alpha}$ defined by 

\begin{eqnarray}
     K_{\parallel \alpha} & =  K_{\parallel \parallel \alpha} +   K_{\parallel \perp \alpha} , \label{eq:kparallel}\\
     K_{\perp \alpha} & =  \frac{1}{2}K_{ \parallel \perp \alpha} +  K_{\perp \perp \alpha} \label{eq:kperp},
\end{eqnarray}
with

\begin{eqnarray}
     K_{\parallel \parallel \alpha} & = \frac{1}{2}m_\alpha^2 \int d W f_\alpha v_\parallel^4 \nonumber \\ 
     & = \tau_\alpha^2 \left( \sqrt{6}\mathcal{N}_\alpha^{40}+  3 \sqrt{2} \mathcal{N}_\alpha^{20}+ \frac{3}{2}   \mathcal{N}_\alpha^{00}\right), \\
     K_{\parallel \perp \alpha} & =  m_\alpha  \int d W f_\alpha v_\parallel^2 \mu B \nonumber \\
     & =  \tau_\alpha^2 \left( \sqrt{2} \mathcal{N}_\alpha^{20} + \mathcal{N}_\alpha^{00} - \sqrt{2}\mathcal{N}_\alpha^{21}- \mathcal{N}_\alpha^{01}\right),  \\
     K_{\perp \perp \alpha} &= \int d W f_\alpha ( \mu B)^2  \nonumber \\
     & = 2 \tau_\alpha^2  \left(\mathcal{N}_\alpha^{02} + \mathcal{N}_\alpha^{00} - 2 \mathcal{N}_\alpha^{01} \right).
\end{eqnarray}
Here, the spectral expansion given in \eref{eq:faspectral} is used to performed explicitly the velocity-space integrals.

From \eref{eq:gammaD} and \eref{eq:qD}, we observe that, unlike the electrostatic $\bm{E} \times \bm{B}$ particle and heat fluxes, the diamagnetic fluxes depend on higher-order spectral coefficients due to the velocity dependence of $\bm{V}_{D\alpha}$ \eref{eq:vdia}. This suggests that the diamagnetic fluxes are more sensitive to kinetic features. To capture these kinetic effects, a spectral resolution of at least $(5,3)$ is necessary. Hence, the diamagnetic heat fluxes may not be accurately described using a resolution of, e.g., $(4,2)$ since the contributions for higher-order spectral coefficients are artificially suppressed in quantities such as $K_{\parallel \parallel \alpha}$ and $K_{\perp \perp \alpha}$. Therefore, it is worth investigating the accuracy of the fluid limit of the diamagnetic fluxes. Assuming a local Maxwellian distribution function with isotropic pressure, the expressions \eref{eq:gammaD} and \eref{eq:qD} reduce to the fluid diamagnetic fluxes, denoted by $\Gamma_{D \alpha}^{\mathrm{Fluid}}$ and $Q_{D \alpha}^{\mathrm{Fluid}}$. These are given by

\begin{eqnarray} \label{eq:gammadiafluid}
    \Gamma_{D \alpha}^{\mathrm{Fluid}} & = \frac{c P_\alpha}{q_\alpha B} \left( \nabla \times \bm{b} + \frac{1}{B} \bm{b} \times \nabla B \right) \cdot \nabla r, \\
    Q_{D \alpha}^{\mathrm{Fluid}} & = \frac{5}{2} \frac{c P_\alpha T_\alpha}{q_\alpha B} \left( \nabla \times \bm{b} + \frac{1}{B} \bm{b} \times \nabla B \right) \cdot \nabla r. \label{eq:qdiafluid}
\end{eqnarray}
The fluid diamagnetic fluxes, given in \eref{eq:gammadiafluid} and \eref{eq:qdiafluid}, do not resemble the classical diamagnetic heat flux typically found in drift-reduced Braginskii models (see, e.g., \cite{zholobenko2024}), which is proportional to $5/2 P_\alpha \bm{U}_{\alpha}^* \cdot \nabla r$, where $\bm{U}_{\alpha}^* = c\bm{B} \times \nabla P_\alpha / (q_\alpha n_\alpha B^2)$ is the fluid diamagnetic velocity. However, it can be shown that their difference is divergence-free using the identity

\begin{equation}
  \nabla \cdot \left[ \frac{P_{\alpha}}{B} \left( \nabla \times \bm{b} + \frac{\bm{b} \times \nabla B}{B} \right) \right] = \nabla \cdot \left( - \frac{\nabla P_\alpha \times \bm{b}}{B} \right).
\end{equation}
As a result, both fluxes yield the same transport across a flux-surface. 

\begin{figure}
    \centering
\includegraphics[scale=0.35]{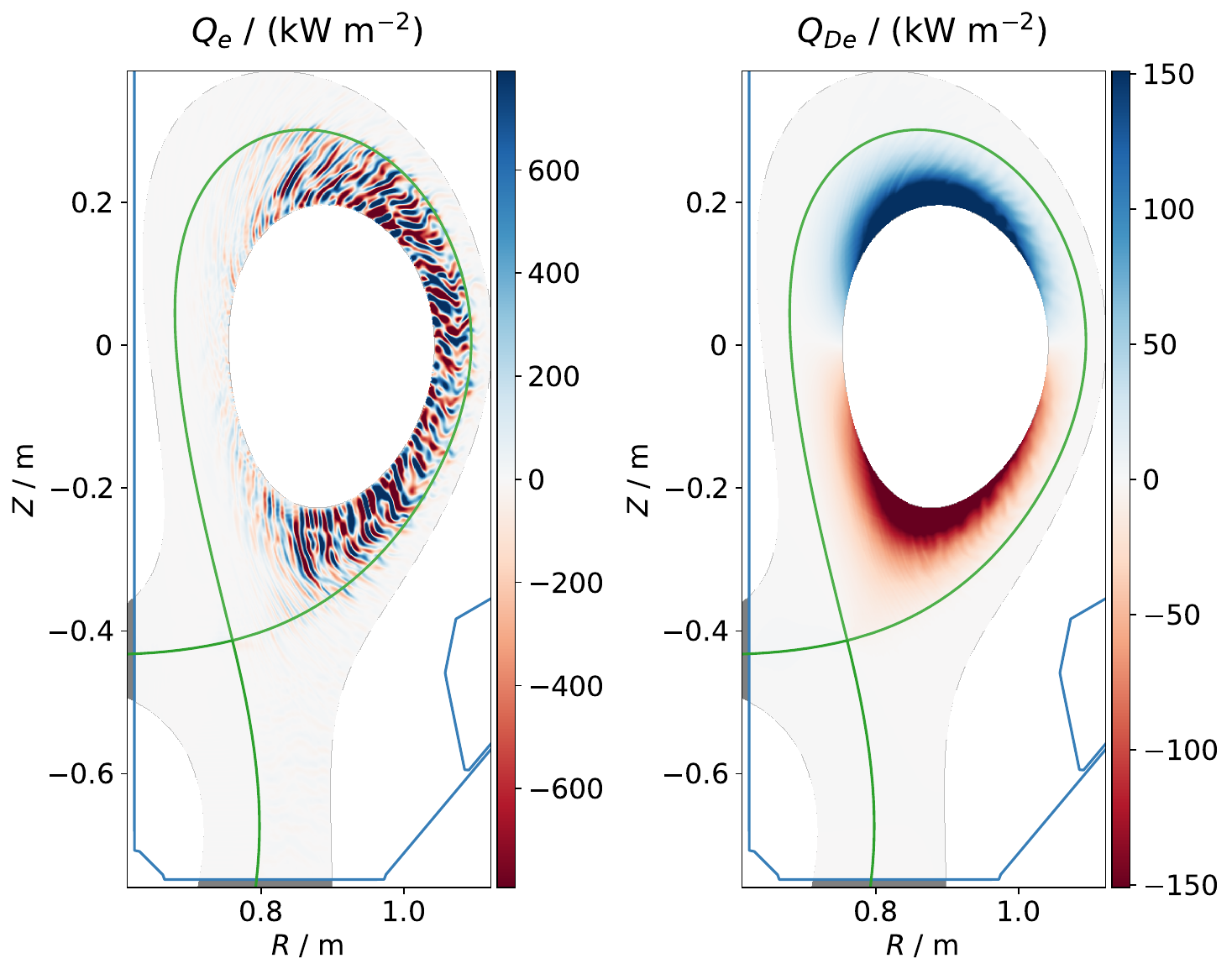}
     \caption{Snapshots of the electron radial $\bm{E} \times \bm{B}$ heat flux $Q_e$ (left) and the diamagnetic heat flux $Q_{De}$ (right) in a single poloidal plane from the $(8,4)$ spectral simulation.}
    \label{fig:exbdiamfluxes}
\end{figure}

\Fref{fig:exbdiamfluxes} shows snapshots of the radial components of the electron heat flux $Q_e$, associated with the $\bm{E} \times \bm{B}$ velocity, \eref{eq:qalpha}, and the diamagnetic heat flux $Q_{De}$ given in \eref{eq:qD}. The heat fluxes exhibit distinct spatial structures. While $Q_e$ features clear turbulent structures on the low-field side due to the ballooning characteristics of TEM, $Q_{De}$ displays a dominant up-down asymmetry pronounced in the edge region and subdominant filamentary turbulent structures extending toward the separatrix. The same observations can be made with the particle fluxes, but not shown here. The up-down asymmetry in $Q_{De}$ arises because of the poloidal inhomogeneity of the magnetic field, which enters in the geometric factors in \eref{eq:qD} and \eref{eq:qD}. We note that the ion diamagnetic flux exhibits a reversed up-down asymmetry compared to $Q_{De}$ because $Q_{D\alpha}$ is proportional to the electric charge.

\begin{figure}
    \centering
    \includegraphics[scale=0.42]{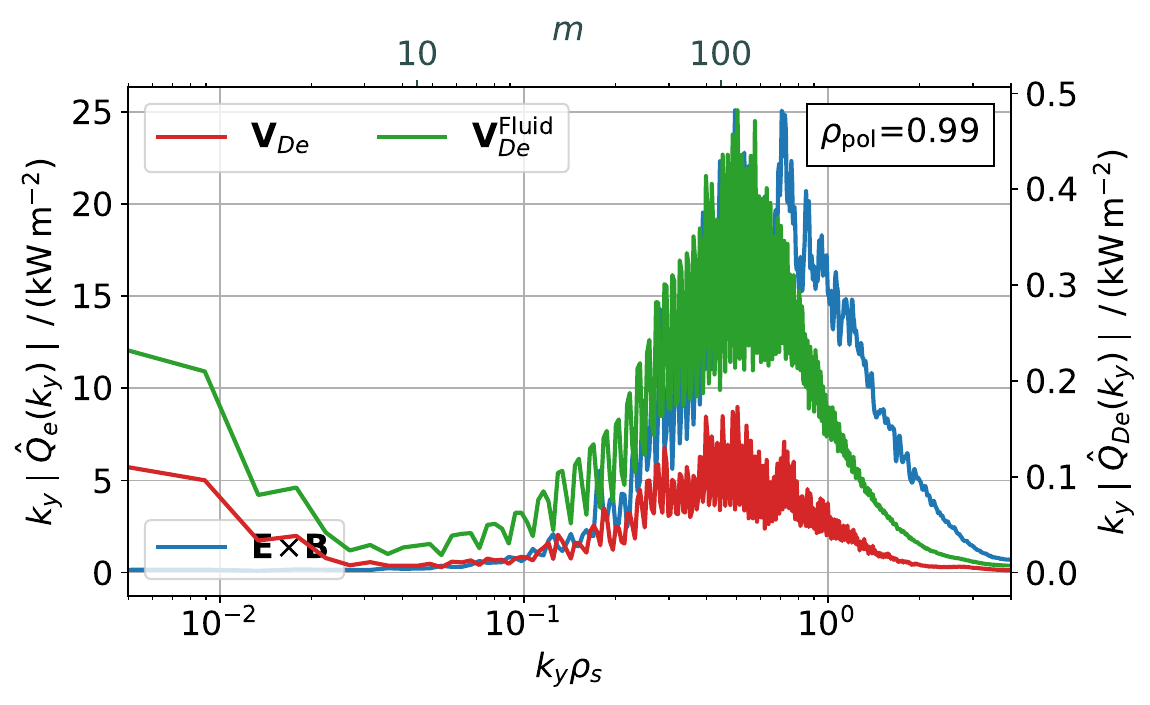}
     \caption{Spectra of the electron $\bm E \times \bm B$ turbulent $Q_e$ (blue line and left y-axis) and diamagnetic $Q_{De}$ (associated with $\bm V_{D e}$) (red line and right y-axis) heat fluxes evaluated on the $\rho_{\mathrm{pol}} = 0.99$ flux-surface. For comparison, we also show the fluid (associated with a local Maxwellian distribution function) diamagnetic flux associated with $\bm V_{D e}^{\mathrm{Fluid}}$ in green (right y-axis). The spectra are computed on a single poloidal plane and averaged over $0.1$~ms. Similar spectra are obtained for the ions, but not shown.}
    \label{fig:fourierfluxexbdiam}
\end{figure}

To further analyze the spatial structures of $\Gamma_{D\alpha}$ and $Q_{D\alpha}$, we perform a Fourier analysis near the separatrix and compare the spectra with those of the $\bm{E} \times \bm{B}$ particle and heat fluxes diagnosed in \sref{subsec:fourierspectrafluxes}. The results, shown in \fref{fig:fourierfluxexbdiam}, reveal that the Fourier components of the diamagnetic fluxes are at least two orders of magnitude smaller than those of the turbulent $\bm{E} \times \bm{B}$ fluxes. The $\bm{E} \times \bm{B}$ spectra exhibit a pronounced peak near $k_y \rho_s \sim 0.5$, driven by turbulence, while the diamagnetic spectra show an increase of the long wavelength components in addition to a peak (of similar amplitude) near $k_y \rho_s \sim 0.5$ and another at longer wavelengths. The peak at $k_y \rho_s \sim 0.5$ is attributed to turbulence-induced diamagnetic fluxes, while the long wavelength components stem from neoclassical physics \cite{helander2005} as depicted in \fref{fig:exbdiamfluxes}. We note that a comprehensive study of the neoclassical contribution to the diamagnetic fluxes is left for future work.

\Fref{fig:fourierfluxexbdiam} also shows the Fourier spectra of the electron fluid diamagnetic heat flux given in \eref{eq:qdiafluid}. It can be observed that the fluid approximation systematically overestimates the diamagnetic heat fluxes by approximately a factor of $3$. A similar trend is observed for the particle flux but not shown here. This discrepancy highlights the importance of kinetic corrections (e.g., $K_{\parallel \alpha}$ and $K_{\perp \alpha}$ in \eref{eq:kparallel} and \eref{eq:kperp} as well as pressure anisotropy) for accurately predicting their amplitude.

Integrating the diamagnetic fluxes across the separatrix show that they contribute to approximately $1.5 \times 10^{20}$~1/s to the radial particle transport, compared to $2.5 \times 10^{24}$~1/s for the turbulent $\bm{E} \times \bm{B}$ particle flux. Similarly, the diamagnetic contribution to the total power $P_{\mathrm{sep}}$, defined in \eref{eq:psep}, is estimated to be $1.7$~kW for electrons and less than $0.1$~kW for ions, which is smaller by at least one order of magnitude than the $\bm{E} \times \bm{B}$ heat flux. This confirms that the radial particle and heat transport are dominated by the $\bm{E} \times \bm{B}$ velocity, justifying the absence of $Q_{D\alpha}$ in \eref{eq:qalpha}. Although the diamagnetic fluxes are negligible in the present L-mode discharge, it has been found in global drift-reduced Braginskii simulations \cite{zholobenko2024} that they play a significant role in high-performance scenarios. In this context, the discrepancies between the fluid approximation and the exact form of the diamagnetic fluxes highlight the need for a kinetic description in high-performance conditions.

\section{Qualitative Comparison with Braginskii Model} \label{sec:comparisonbraginskii}

A comparison with a lower-fidelity approach is valuable for assessing and contrast the impact of neglecting TEMs on turbulence and transport. Since trapped particles cannot be easily excluded in the spectral approach, we instead perform a qualitative comparison with the drift-reduced Braginskii fluid code \texttt{GRILLIX} \cite{stegmeir2019,zholobenko2023}, which neglects trapped particle dynamics. The \texttt{GRILLIX} code solves drift-reduced fluid equations for density, parallel velocities, and temperatures, along with a vorticity equation and Ampere’s law for electrostatic and magnetic vector potentials (see \cite{stegmeir2019} for details). Compared to the original TCV-X21 validation study \cite{oliveira2022}, magnetic flutter effects are now included, which have been shown to significantly reduce turbulent transport, even in L-mode conditions \cite{zhang2024}. The electron and ion parallel heat conductivities are modelled using the Braginskii closure with heat-flux limiters set to $\alpha_e = \alpha_i = 1.0$. These values are chosen to reproduce the Landau-fluid closure \cite{pitzal2023}. We emphasize that adopting lower values for the heat-flux limiters can artificially enhance the edge transport and enables, e.g., realistic simulations of the SOL, where TEMs are expected to be negligible and heat flux limiters become irrelevant, as the Braginskii closure is applicable. However, this artificially increases the edge transport. Therefore, we consider $\alpha_e = \alpha_i = 1.0$ for the present work to study the impact of the absence of TEMs in the edge. 

The numerical setup follows \sref{sec:modellingsetup}, with adjustments due to fundamental differences between the GK and Braginskii models. At the inner boundary, density and temperature profiles are matched to \texttt{GENEX} at $\rho_{\mathrm{pol}} = 0.75$ using adaptive sources, while homogeneous Neumann conditions are applied for all dynamically evolved fluid quantities. Sheath boundary conditions are used close to the divertor targets (see \cite{stegmeir2019} for details), unlike in GENE-X. Thus, we restrict the comparison herein to the plasma edge. Unlike in \cite{oliveira2022}, where sources were added near the separatrix to match experimental values of $n_e$ and $T_e$, the present study constrains density and temperature at $\rho_{\mathrm{pol}} = 0.75$, allowing free variation near the separatrix.

\begin{figure}
    \centering
    \includegraphics[scale=0.5]{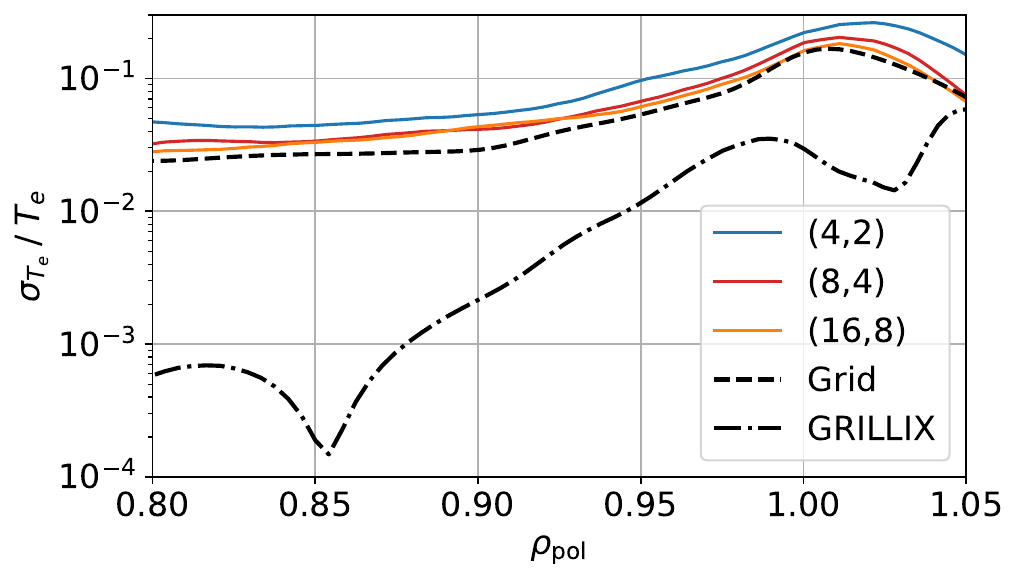}
    \includegraphics[scale=0.5]{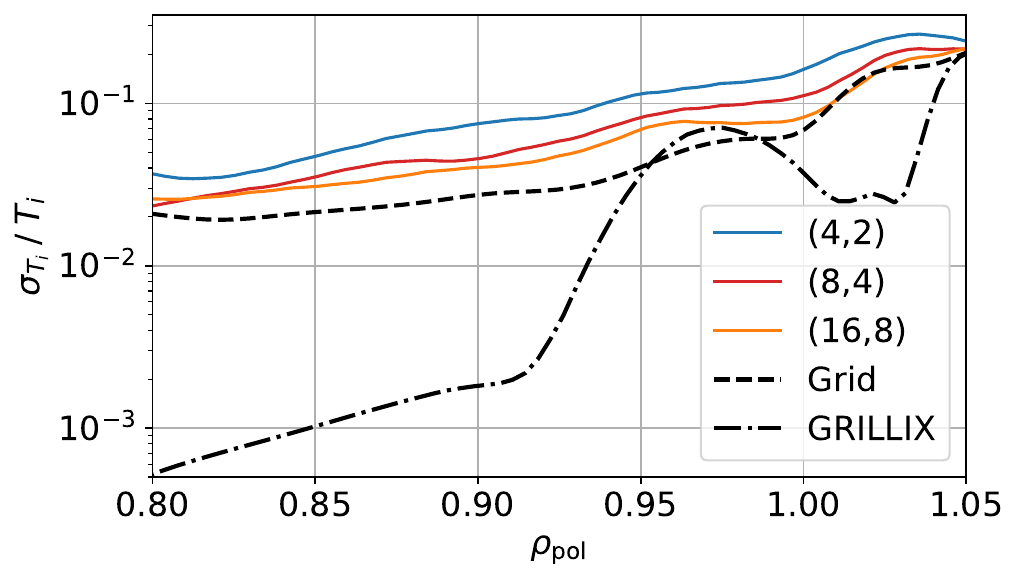}
    \caption{Relative electrons (left) and ions (right) temperature fluctuation amplitudes, $\sigma_{T_\alpha} / T_\alpha$, measured at the OMP. The colored lines represent the results from spectral simulations, while the dashed black lines correspond to grid-based simulations, and the dotted black line shows the \texttt{GRILLIX} results. The standard deviations, $\sigma_{T_\alpha}$, are calculated from toroidal and time averages (over $0.1$~ms) in steady state.}
    \label{fig:ompfluc}
\end{figure}

\Fref{fig:ompfluc} compares the relative temperature fluctuation amplitudes at the OMP for electrons and ions, as obtained from GK  (both spectral and grid-based) simulations and from \texttt{GRILLIX} Braginskii fluid simulations. The spectral and grid-based GK results exhibit comparable fluctuation levels, although the $(4,2)$ model shows notably higher amplitudes. In contrast, the \texttt{GRILLIX} simulation yields negligible temperature fluctuations for $\rho_{\mathrm{pol}} \lesssim 0.95$. More specifically, electron temperature fluctuations remain below 5\% in \texttt{GRILLIX}, while in GK simulations they increase radially, reaching up to 15\% due to TEM activity. Ion temperature fluctuations in \texttt{GRILLIX} peak at $\sigma_{T_i} / T_i \sim 0.07$ near $\rho_{\mathrm{pol}} = 0.98$, where small amplitude fluctuations are driven by steep local $T_i$ gradients. Additionally, \texttt{GRILLIX} obtains lower separatrix temperatures and densities (not shown). Near $\rho_{\mathrm{pol}} = 0.98$, strong poloidal and sheared flows emerge in \texttt{GRILLIX}, driven by ion pressure gradients. These flows generate a sheared radial electric field with negative values around $-15$~kV/m, in contrast to the GK simulations where $E_r$ nearly vanishes at this location (see \fref{fig:omper}).

The turbulence regime and transport characteristics in \texttt{GRILLIX} differ substantially from those observed in \sref{sec:turbulencecharacteristics} and \sref{sec:turbulenttranapsortanalysis}. A temporal Fourier analysis (analogous to \sref{sec:turbulencecharacteristics}) identifies toroidal ITG modes as the dominant instability, contrasting with the TEM-driven turbulence in the GK simulations. Consequently, the ion heat transport is found to be larger than the electron transport. Furthermore, the power across the separatrix in \texttt{GRILLIX} is an order of magnitude lower than in \texttt{GENEX} and experimental measurements, a consequence of the weak turbulence observed in the fluid model. Given that TEMs carry most of the heat flux in experimental devices such as TCV, these results suggest that the use of a Braginskii-like fluid model can lead to unrealistic transport predictions in certain conditions \cite{ulbl2025}.

We note that the Braginskii model implemented in \texttt{GRILLIX} includes the lowest-order FLR effect, resulting in a diamagnetic correction to the vorticity equation, while the associated term is neglected in the \texttt{GENE-X} model due to the long-wavelength approximation adopted in the former. Although this FLR effect can stabilize ion-scale microinstabilities such as ITG, it is not expected to significantly affect the qualitative comparison in this section, which is primarily driven by TEM-turbulence. Future work aims at including these effects in \texttt{GENE-X}.

Despite the fact that the present findings are specific to the TCV-X21 case, they point to possible improvements in Braginskii-like models. The strong deviations observed in the fluid predictions and the relatively good overall agreement achieved by the  $(4,2)$ spectral simulation with grid-based GK results suggest that pressure anisotropy is crucial for capturing trapped particle physics. Incorporating pressure anisotropy in Braginskii-like models could improve their ability to approximate the kinetic response of trapped particles, providing an extension towards low collisionality regime.

\section{Conclusions}
\label{sec:conclusions}
This work presents a comprehensive physical interpretation and analysis of turbulence and transport using the full-$f$ GK spectral approach recently implemented in the \texttt{GENE-X} code \cite{frei2024}. Through extensive comparison with previous grid-based GK simulations \cite{ulbl2023}, we have demonstrated that the spectral method reproduces, with high fidelity, all key turbulence and transport features (including OMP profiles and gradients, radial electric fields, and edge turbulence characteristics) previously identified in the TEM-dominated TCV-X21 scenario.

The central focus of the first part of the present work was the self-consistent evolution of the slowly varying and long-wavelength radial electric field $E_r$ within the full-$f$ GK formalism, which complements the previous study \cite{ulbl2023}. In particular, a detailed investigation of the radial force balance showed that the full-$f$ approach consistently captures the interplay between $E_r$, poloidal flows, and toroidal flows, maintaining accurate force balance. A systematic decomposition of $E_r$ revealed that its edge structure is dominated by the competition between the ion pressure gradient and toroidal rotation, while poloidal flows become increasingly important near the separatrix. In the SOL, the amplitude of $E_r$ is primarily set by parallel electron dynamics. These results validate the ability of the \texttt{GENE-X} model (both grid-based and spectral) to accurately resolve the dynamics of long-wavelength electric fields and flows, a fundamental requirement for full-$f$ GK modeling, as they play a central role in regulating turbulent transport.

The turbulent characteristics and the resulting transport predicted by the spectral approach were then analyzed using comprehensive temporal and spatial Fourier diagnostics. The results demonstrate that the spectral simulations accurately reproduce all key features (spatial and temporal spectra and phase-shifts) of TEM-driven turbulence and transport, showing excellent agreement with grid-based simulations \cite{ulbl2023}. Furthermore, these results exhibit minimal sensitivity to spectral resolution, consistent with earlier linear convergence studies \cite{frei2024}. A detailed Fourier analysis of the fluxes (including their phase shifts) confirms the dominance of electrons in the radial heat flux, a signature of TEM-driven turbulence. Furthermore, the predicted power crossing the separatrix agrees with experimental measurements within $10\%$, matching grid-based predictions, while the SOL fall-off length is found to be qualitatively consistent with previous predictions. Finally, under the present L-mode conditions, the diamagnetic (neoclassical) fluxes are found to be negligible.

To emphasize the key role of TEM-driven transport in the TCV-X21 scenario, a qualitative comparison was performed with Braginskii-like fluid simulations using the \texttt{GRILLIX} code. In the absence of TEMs, the fluid simulations exhibit steep OMP profiles, strong radial electric fields, reduced fluctuation amplitudes predominantly dominated by ITG-turbulence. These observations stand in strong contrast to the GK simulation results. Given the overall good agreement between the lowest spectral resolution case, i.e., $(4,2)$, and the grid-based GK results, we suggest that incorporating pressure anisotropy into Braginskii-like fluid models could provide a significant improvement in their physics fidelity under edge conditions. This low-collisionality extension may help capture trapped particle effects and enhance the accuracy of turbulence and transport predictions in fluid-based models.

Finally, while a spectral resolution of $(6,4)$ was previously shown in \cite{frei2024} to be sufficient for capturing the OMP profiles, the present study demonstrates that turbulence and the associated transport are also accurately resolved with at least $(8,4)$ spectral coefficients. This work has primarily focused on the edge region, and a more detailed investigation of the SOL remains necessary, particularly given that the current \texttt{GENE-X} model employs simplified boundary conditions. Furthermore, it remains an open question how the present findings extrapolate to high-performance regimes and reactor-relevant conditions, although preliminary results suggest that the observed trends remain robust. Therefore, by combining the computational efficiency established in \cite{frei2024} with the high physical fidelity demonstrated in the present work, the full-$f$ GK spectral approach in \texttt{GENE-X} offers a promising path toward first-principles and predictive simulations of edge and SOL turbulence for future reactor design.

\section*{Acknowledgement}

The authors would like to thank G. Dif-Pradalier for fruitful discussion about the radial force balance.

This work has been carried out within the framework of the EUROfusion Consortium, funded by the European Union via the Euratom Research and Training Programme (Grant Agreement No 101052200 — EUROfusion). Views and opinions expressed are however those of the author(s) only and do not necessarily reflect those of the European Union or the European Commission. Neither the European Union nor the European Commission can be held responsible for them.

\appendix

\section{Linear Frequency Estimates} \label{appendix:linearfrequency}

This appendix reports the analytical expressions of the linear frequency estimates used to identify the underlying modes driving turbulence in \sref{subsec:temporalfourier}.

One of the most fundamental modes driven by density gradients is the electron drift wave (eDW). The linear frequency of the eDW, denoted by $\omega_e^*$, is given by

\begin{equation} \label{eq:wdw}
    \omega_e^* = - \frac{T_e}{e B} \nabla_r \ln n \, k_y,
\end{equation}
where $\nabla r \cdot \nabla = \nabla_r$. For $\nabla_r \ln n < 0$, $\omega_e^*$ is positive, indicating that eDW propagates in the electron diamagnetic direction in our convention.

A microinstability of interest is the ion temperature gradient (ITG) mode. This instability can be destabilized by the presence of finite ion temperature gradients through parallel motion and magnetic gradient drift resonance effects. In the case of ITG destabilized by parallel motion only (slab branch), the linear frequency, $\omega_{i}^*$, is given by \cite{brunner1997,frei2022}

\begin{equation} \label{eq:wsitg}
    \omega_{i}^* = \frac{k_y}{q_i B} \nabla_r T_i.
\end{equation}
We note that $\omega_{i}^*$ is negative, indicating that the ITG mode propagates in the ion diamagnetic direction.

Finally, another microinstability that can drive substantial transport in the edge is the trapped electron mode (TEM). Two branches of TEM exist, depending on plasma collisionality and density gradient. At low collisionality, the collisionless TEM (cTEM) can occur \cite{catto1978,ernst2004,dannert2005}, while at finite collisionality, the dissipative TEM (dTEM) can emerge \cite{chen1977,connor2006,zhao2017}. In the collisionless limit, an analytical estimate for the linear frequency of cTEM, $\omega_{\mathrm{cTEM}}$, can be obtained by neglecting ion parallel motion and magnetic drifts. Assuming the mode frequency is much larger than the electron precessional drift frequency, $\omega_{\varphi e} \simeq L_N \omega_e^* / R$, the linear estimate for $\omega_{\mathrm{cTEM}}$ is

\begin{equation}\label{eq:wctem}
    \omega_{\mathrm{cTEM}} \simeq \frac{1}{2 (1 - f_t)} \left[ \omega_e^* (1 - f_t) + \frac{3}{2} f_t \omega_{\varphi e} \right],
\end{equation}
where $ f_t = \sqrt{1 - B_{\mathrm{min}} / B_{\mathrm{max}}} $ is the electron trapping fraction.

A linear estimate of the dTEM frequency, $\omega_{\mathrm{dTEM}}$, can also be derived in the presence of steep density gradients, an assumption applicable in both internal transport barriers and edge region. In this case, $\omega_{\mathrm{dTEM}}$ is given by \cite{connor2006}

\begin{equation} \label{eq:wdtem}
    \omega_{\mathrm{dTEM}} \simeq \omega_e^* \left( \frac{I_0(b) e^{-b} + \eta_i b \left(I_1(b) - I_0(b)\right) e^{-b}}{1 + T_e / T_i \left(1 - I_0(b)e^{-b}\right)} \right),
\end{equation}
where $ b = (k_y \rho_i)^2 $, and $ I_0 $ and $ I_1 $ are the modified Bessel functions of the first kind of order zero and one, respectively. Unlike \eref{eq:wctem}, the linear frequency of dTEM has a non-trivial dependence on the wavenumber $ k_y $ which follows $k_y^{-2}$ at short perpendicular wavelengths. We note that dTEM primarily drives transport at small perpendicular wavelengths, as strong stabilization occurs at long wavelengths \cite{connor2006}.

% References
% Load Bibliography file 
\section*{References}
\bibliography{library}

\providecommand{\newblock}{}
\begin{thebibliography}{10}
\expandafter\ifx\csname url\endcsname\relax
  \def\url#1{{\tt #1}}\fi
\expandafter\ifx\csname urlprefix\endcsname\relax\def\urlprefix{URL }\fi
\providecommand{\eprint}[2][]{\url{#2}}
% Bibliography created with iopart-num v2.1
% /biblio/bibtex/contrib/iopart-num

\bibitem{ikeda2007}
Ikeda K 2007 {\em Nuclear Fusion\/} {\bf 47} E01

\bibitem{zohm2013}
Zohm H, Angioni C, Fable E, Federici G, Gantenbein G, Hartmann T, Lackner K,
  Poli E, Porte L, Sauter O {\em et~al.\/} 2013 {\em Nuclear Fusion\/} {\bf 53}
  073019

\bibitem{litaudon2022}
Litaudon X, Jenko F, Borba D, Borodin D~V, Braams B~J, Brezinsek S, Calvo I,
  Coelho R, Donn{\'e} A~J~H, Embr{\'e}us O {\em et~al.\/} 2022 {\em Plasma
  Physics and Controlled Fusion\/} {\bf 64} 034005

\bibitem{brizard2007}
Brizard A~J and Hahm T~S 2007 {\em Reviews of modern physics\/} {\bf 79} 421

\bibitem{zeiler1997}
Zeiler A, Drake J~F and Rogers B 1997 {\em Physics of Plasmas\/} {\bf 4} 2134

\bibitem{frei2020}
Frei B~J, Jorge R and Ricci P 2020 {\em Journal of Plasma Physics\/} {\bf 86}
  905860205

\bibitem{ulbl2023}
Ulbl P, Body T, Zholobenko W, Stegmeir A, Pfennig J and Jenko F 2023 {\em
  Physics of Plasmas\/} {\bf 30} 107986

\bibitem{frei2022}
Frei B~J, Hoffmann A~C~D and Ricci P 2022 {\em Journal of Plasma Physics\/}
  {\bf 88} 905880304

\bibitem{mandell2020}
Mandell N~R, Hakim A, Hammett G~W and Francisquez M 2020 {\em Journal of Plasma
  Physics\/} {\bf 86} 905860109

\bibitem{michels2021}
Michels D, Stegmeir A, Ulbl P, Jarema D and Jenko F 2021 {\em Comput. Phys.
  Commun.\/} {\bf 264} 107986

\bibitem{hager2022}
Hager R, Ku S, Sharma A~Y, Chang C~S, Churchill R~M and Scheinberg A 2022 {\em
  Physics of Plasmas\/} {\bf 29}

\bibitem{bottino2025}
Bottino A, Stier A, Boesl M, Hayward-Schneider T, Bergmann A, Coster D, Brunner
  S, Di~Giannatale G and Villard L 2025 {\em Plasma Physics and Controlled
  Fusion\/} {\bf 67} 025008

\bibitem{grandgirard2007}
Grandgirard V, Sarazin Y, Angelino P, Bottino A, Crouseilles N, Darmet G,
  Dif-Pradalier G, Garbet X, Ghendrih P, Jolliet S {\em et~al.\/} 2007 {\em
  Plasma Physics and Controlled Fusion\/} {\bf 49} B173

\bibitem{tamain2016}
Tamain P, Bufferand H, Ciraolo G, Colin C, Galassi D, Ghendrih P, Schwander F
  and Serre E 2016 {\em Journal of Computational Physics\/} {\bf 321} 606

\bibitem{stegmeir2019}
Stegmeir A, Ross A, Body T, Francisquez M, Zholobenko W, Coster D, Maj O, Manz
  P, Jenko F, Rogers B~N {\em et~al.\/} 2019 {\em Physics of Plasmas\/} {\bf
  26} 052517

\bibitem{giacomin2021}
Giacomin M, Ricci P, Coroado A, Fourestey G, Galassi D, Lanti E, Mancini D,
  Richart N, Stenger L and Varini N 2021 {\em J. Comput. Phys.\/} {\bf 463}
  111294

\bibitem{pitzal2023}
Pitzal C, Stegmeir A, Zholobenko W, Zhang K and Jenko F 2023 {\em Physics of
  Plasmas\/} {\bf 30}

\bibitem{frei2024}
Frei B~J, Ulbl P, Trilaksono J and Jenko F 2024 {\em arXiv preprint
  arXiv:2411.09232\/}

\bibitem{oliveira2022}
Oliveira D~S, Body T, Galassi D, Theiler C, Laribi E, Tamain P, Stegmeir A,
  Giacomin M, Zholobenko W, Ricci P {\em et~al.\/} 2022 {\em Nuclear Fusion\/}
  {\bf 62} 096001

\bibitem{dif2009}
Dif-Pradalier G, Grandgirard V, Sarazin Y, Garbet X and Ghendrih P 2009 {\em
  Physical Review Letters\/} {\bf 103} 065002

\bibitem{zholobenko2021}
Zholobenko W, Body T, Manz P, Stegmeir A, Zhu B, Griener M, Conway G~D, Coster
  D, Jenko F, Team A~U {\em et~al.\/} 2021 {\em Plasma Physics and Controlled
  Fusion\/} {\bf 63} 034001

\bibitem{scott2010}
Scott B and Smirnov J 2010 {\em Physics of Plasmas\/} {\bf 17}

\bibitem{dougherty1964}
Dougherty J~P 1964 {\em The Physics of Fluids\/} {\bf 7} 1788

\bibitem{ulbl2022}
Ulbl P, Michels D and Jenko F 2022 {\em Contributions to Plasma Physics\/} {\bf
  62} e202100180

\bibitem{michels2022}
Michels D, Ulbl P, Zholobenko W, Body T, Stegmeir A, Eich T, Griener M, Conway
  G~D, Jenko F, Team A~U {\em et~al.\/} 2022 {\em Physics of Plasmas\/} {\bf
  29}

\bibitem{gradshteyn2014}
Gradshteyn I~S and Ryzhik I~M 2014 {\em Table of integrals, series, and
  products\/} (Academic press)

\bibitem{frei2023}
Frei B~J, Hoffmann A~C~D, Ricci P, Brunner S and Tecchioll Z 2023 {\em Journal
  Of Plasma Physics\/} {\bf 89} 905890414

\bibitem{tcvx21zenodo}
Ulbl P, Body T, Zholobenko W, Stegmeir A, Pfennig J and Jenko F 2023
  {TCV-X21-GENEX}: influence of collisions on the validation of global
  gyrokinetic simulations \urlprefix\url{https://zenodo.org/record/7894731}

\bibitem{ulbl2023phd}
Ulbl P 2023 {\em Collision Models for Gyrokinetic Simulations of Edge
  Turbulence in Fusion Plasmas\/} Ph.D. thesis Technische Universit{\"a}t
  M{\"u}nchen

\bibitem{hazeltine2013}
Hazeltine R~D and Meiss J~D 2013 {\em Plasma confinement\/} (Courier
  Corporation)

\bibitem{kim1991}
Kim Y~B, Diamond P~H and Groebner R~J 1991 {\em Physics of Fluids B: Plasma
  Physics\/} {\bf 3} 2050

\bibitem{plank2023}
Plank U, Brida D, Conway G, Happel T, Hubbard A, P{\"u}tterich T, Angioni C,
  Cavedon M, Dux R, Eich T {\em et~al.\/} 2023 {\em Physics of Plasmas\/} {\bf
  30}

\bibitem{connor2006}
Connor J~W, Hastie R~J and Helander P 2006 {\em Plasma physics and controlled
  fusion\/} {\bf 48} 885

\bibitem{zholobenko2023}
Zholobenko W, Pfennig J, Stegmeir A, Body T, Ulbl P, Jenko F, Team A~U {\em
  et~al.\/} 2023 {\em Nuclear Materials and Energy\/} {\bf 34} 101351

\bibitem{dannert2005}
Dannert T and Jenko F 2005 {\em Physics of Plasmas\/} {\bf 12}

\bibitem{eich2013}
Eich T, Leonard A~W, Pitts R~A, Fundamenski W, Goldston R~J, Gray T~K, Herrmann
  A, Kirk A, Kallenbach A, Kardaun O {\em et~al.\/} 2013 {\em Nuclear fusion\/}
  {\bf 53} 093031

\bibitem{chang2017}
Chang C~S, Ku S, Loarte A, Parail V, Koechl F, Romanelli M, Maingi R, Ahn J~W,
  Gray T, Hughes J {\em et~al.\/} 2017 {\em Nuclear Fusion\/} {\bf 57} 116023

\bibitem{zholobenko2024}
Zholobenko W, Zhang K, Stegmeir A, Pfennig J, Eder K, Pitzal C, Ulbl P, Griener
  M, Radovanovic L, Plank U {\em et~al.\/} 2024 {\em Nuclear Fusion\/} {\bf 64}
  106066

\bibitem{helander2005}
Helander P and Sigmar D~J 2005 {\em Collisional transport in magnetized
  plasmas\/} vol~4 (Cambridge university press)

\bibitem{zhang2024}
Zhang K, Zholobenko W, Stegmeir A, Eder K and Jenko F 2024 {\em Nuclear
  Fusion\/} {\bf 64} 036016

\bibitem{ulbl2025}
Ulbl P, Stegmeir A, Told D, Merlo G, Zhang K and Jenko F 2025 {\em arXiv
  preprint arXiv:2504.00475\/}

\bibitem{brunner1997}
Brunner S 1997 {\em Global approach to the spectral problem of
  microinstabilities in tokamak plasmas using a gyrokinetic model\/} Ph.D.
  thesis EPFL Lausanne
  \urlprefix\url{https://infoscience.epfl.ch/handle/20.500.14299/210187}

\bibitem{catto1978}
Catto P~J and Tsang K~T 1978 {\em The Physics of Fluids\/} {\bf 21} 1381

\bibitem{ernst2004}
Ernst D~R, Bonoli P~T, Catto P~J, Dorland W, Fiore C~L, Granetz R~S, Greenwald
  M, Hubbard A~E, Porkolab M, Redi M~H {\em et~al.\/} 2004 {\em Physics of
  Plasmas\/} {\bf 11} 2637

\bibitem{chen1977}
Chen L, Berger R~L, Lominadze J~G, Rosenbluth M~N and Rutherford P~H 1977 {\em
  Physical Review Letters\/} {\bf 39} 754

\bibitem{zhao2017}
Zhao C, Zhang T and Xiao Y 2017 {\em Physics of Plasmas\/} {\bf 24}

\end{thebibliography}

\end{document}